\documentclass[12pt,preprint,nofootinbib,noshowkeys,noshowpacs,longbibliography,superscriptaddress,tightenlines,floatfix,longbibliography]{revtex4-2}
\usepackage{amsmath,amsfonts,amssymb}
\usepackage{graphics,graphicx}
\usepackage[inline]{enumitem}
\usepackage{mathrsfs}

\usepackage{mathrsfs}
\usepackage{bm}
\usepackage{color}
\definecolor{darkblue}{RGB}{0,0,196}
\definecolor{darkgreen}{RGB}{0,120,0}

\usepackage{stackengine}

\renewcommand\S{\mathcal S}
\newcommand\Sstress{{\mathcal S}}

\newcommand\el{e}
\newcommand\Lh{{h}}
\def\half{\tfrac{1}{2}}
\def\quarter{\tfrac{1}{4}}

\newcommand{\ubeta}{{\underline{\smash{\beta}}}}

\def\bb{{(b)}}

\def\pr{{\mathcal P}}
\def\er{{\rm e}}

\def\edense{{\mathcal{E}}}
\def\pdense{M}
\def\dpdense{\delta\kern-0.08em M}
\def\dbeta{\delta\kern-0.08em \beta}

\def\llangle{\left\langle}
\def\rrangle{\right\rangle}

\def\noisekernel{{\kappa}}
\def\Eq#1{eq.~(\ref{#1})}
\def\Eqs#1{eqs.~(\ref{#1})}
\def\eq#1{(\ref{#1})}
\def\app#1{App.~\ref{#1}}
\def\Fig#1{Fig.~\ref{#1}}

\def\Sect#1{Sect.~\ref{#1}}

\def\T{\mathcal T}
\def\X{{\mathcal X}}

\def\beq{\begin{equation}}
\def\eeq{\end{equation}}
\def\st{\begin{equation}}
\def\stp{\end{equation}}
\def\ba{\begin{eqnarray}}
\def\ea{\end{eqnarray}}   
\def\DD{{\mathcal L}}

\newcommand\vbeta{{\vec{\beta}}}
\newcommand\vcovD{D}
\newcommand\Chi{\chi}
\newcommand\covD{\nabla}
\newcommand{\ve}[2]{e_{\;\,#1}^{#2}}
\newcommand{\de}[2]{e_{\;\,#2}^{#1}}

\def\dd{{\rm d}}

\def\sp{\phantom{j}}
\def\spm{\phantom{\mu}}
\def\sft{\vec{N}}
\def\tideal{{\mathcal T}}
\def\tvisc{\Pi}
\def\btvisc{\bar{\Pi}}

\newcommand\V{V_0}
\newcommand\dtau{{{\Delta}\tau}}
\newcommand\dt{{{\Delta}t}}

\def\bb1{{(1)}}
\def\bbt{{(2)}}
\def\bbth{{(3)}}

\newcommand{\gammathree}{{}^{(3)}\Gamma}
\usepackage[colorlinks=true,linktocpage=true,linkcolor=darkblue,citecolor=red,urlcolor=darkblue]{hyperref}

\begin{document}
\preprint{}
 
    \title{Stochastic relativistic viscous hydrodynamics from the Metropolis algorithm}
    \author{Jay Bhambure}
    \email{jay.bhambure@stonybrook.edu}	
    \affiliation{Center for Nuclear Theory, Department of Physics and Astronomy, Stony Brook University, Stony Brook, New York, 11794-3800, USA}
    \author{Rajeev Singh}
    \email{rajeev.singh@e-uvt.ro}
    \affiliation{Department of Physics, West University of Timisoara, Bd.~Vasile P\^arvan 4, Timisoara 300223, Romania}
    \author{Derek Teaney}
    \email{derek.teaney@stonybrook.edu}
    \affiliation{Center for Nuclear Theory, Department of Physics and Astronomy, Stony Brook University, Stony Brook, New York, 11794-3800, USA}
	\date{\today} 
	\bigskip
\begin{abstract}
  We propose an algorithm for simulating stochastic relativistic fluid dynamics based on  Metropolis updates. Each step of the algorithm begins with an update based on ideal hydrodynamics. This is followed by proposing random (spatial) momentum transfers between fluid cells, keeping the total energy fixed. These proposals are then accepted or rejected using the change in entropy as a statistical weight. The algorithm reproduces relativistic viscous hydrodynamics in the ``Density Frame",  which is a formulation of viscous hydrodynamics we review and clarify here. This formulation is first order in time and requires no auxiliary dynamical fields such as $\Pi^{\mu\nu}$. The only parameters are the shear and bulk viscosities and the equation of state. By adopting the 3+1 split of general relativity, we extend the Metropolis algorithm to general space-time coordinates,  such as Bjorken coordinates,  which are commonly used to simulate heavy-ion collisions.
\end{abstract}
     
\date{\today}
	
\maketitle
\newpage
\tableofcontents

\section{Introduction}

\subsection{Physical motivation}
Nuclear collisions at the Relativistic Heavy Ion Collider and the Large Hadron
Collider exhibit remarkable collective flows which are well described by
relativistic viscous hydrodynamics without noise~\cite{Heinz:2013th}.  
Current Bayesian fits to the rich phenomenology 
of hydrodynamic correlations have provided increasingly quantitative constraints
on the shear viscosity of QCD and its equation of state~\cite{Bernhard:2019bmu,PhysRevLett.126.202301,JETSCAPE,Heffernan:2023kpm}. 
Strikingly, the shear viscosity to
entropy ratio is measured to be no more than four times a quantum limit of $\hbar/4\pi k_B$, which was suggested by gauge-gravity duality~\cite{Kovtun:2004de}.

In spite of this success, there are  multiple physical motivations 
for developing stochastic hydrodynamics in the relativistic domain.  
Indeed, the strength and importance of the noise is proportional to the 
number of particles in the event $N$, leading to fascinating $1/N$ corrections to
hydrodynamics~\cite{Kovtun:2012rj}.  These corrections lie outside of the usual expansion in the mean-free path to
system size and must be computed from hydrodynamic loops or from
an appropriate set of hydro-kinetic 
equations~\cite{Akamatsu:2016llw,An:2019osr}.
It is important to quantify these $1/N$ corrections for nucleus-nucleus collisions where the number of produced particles is limited.

In fact,  one of the striking findings from the LHC and RHIC is that proton-nucleus and other small colliding systems exhibit collective flow-like correlations, although the number of produced particles in these events is not very large~\cite{PHENIX:2015idk,CMS:2015yux}. The stochastic character of the hydrodynamic motion, which
is a consequence of the finite number of particles,  is paramount for these colliding systems.  

There are other motivations to develop stochastic hydrodynamics. For example,
there are ongoing searches for critical behavior in nucleus-nucleus collisions both
at the LHC and at the fixed target program in STAR~\cite{ALICE3,ALICE3ITS,Du:2024wjm}.
Close to the critical point, modeling stochastic fluctuations is essential to
describing the physics. At high temperature and zero baryon density  there is an
increasing evidence from lattice QCD that QCD is
close to an $O(4)$ critical point,  which describes chiral symmetry
breaking~\cite{Kaczmarek:2020sif}. The experimental evidence for the remnants of $O(4)$
critical dynamics is limited, but there are suggestive hints in the production
of soft  pions~\cite{Grossi:2021gqi}.   At lower temperature  and high baryon density,  there is 
developing evidence from the functional Renormalization Group (fRG)~\cite{Fu:2019hdw} and Taylor
series expansions in lattice QCD~\cite{Borsanyi:2020fev,HotQCD:2018pds,Clarke:2024seq} that an Ising-like critical point
should exist at a temperature and baryon  chemical potential of approximately $T\sim 85$ MeV and $\mu_B/T\sim 6.5$.
This range of temperatures and chemical potentials can be probed by the STAR
fixed target program. However, in this range  of collision energies 
particle production is negligible
and the number of particles in the event is limited by
the number of nucleons in the incoming nuclei,   making this a particularly challenging domain for relativistic fluid dynamics, and a domain where the evolution is decidedly stochastic.

\subsection{The Metropolis algorithm for relativistic fluid dynamics}

Our work is also motivated by recent conceptual and algorithmic advances in 
the study of dissipative relativistic fluids. 
Stochastic processes in thermal equilibrium are naturally
modeled using a Metropolis-like algorithm which automatically respects 
the Fluctuation-Dissipation-Theorem (FDT)~\cite{FoxUhlenbeck,OldMCMC,DUANE1987216,Moore:1998zk,Generic}. 
The approach was used
to simulate the dissipative dynamics of the chiral critical point~\cite{Florio:2021jlx,Florio:2023kmy}. Recently, a pioneering
paper used the Metropolis updates to simulate the real time dynamics of the Ising critical point in QCD~\cite{Chattopadhyay:2024jlh}. 
The methods developed in the current manuscript share the same principles with
the algorithms used in these studies.  First the system is evolved
with a symplectic step of ideal hydrodynamics, and subsequently the conserved charges  
are randomly transferred between fluid cells.
These charge transfers are accepted or rejected using the entropy as a
statistical weight to complete the update.  
Such algorithms are fundamental in statistical mechanics and it  would be surprising,  and profoundly disconcerting,  if they were inapplicable to highly boosted fluids.

The goal of the current work,
which is a continuation of our recent study on relativistic advection diffusion equation~\cite{Basar:2024qxd}, 
is to generalize the Metropolis update algorithm to relativistic flows
and to general coordinate systems. 

First order dissipative fluid dynamics in the relativistic domain as envisioned by the Eckart and  Landau and Lifshitz (LL) is known to have generic instabilities~\cite{Hiscock:1985zz,GAVASSINO2023137854}.  Ultimately these
instabilities arise because the divergence of the viscous stress tensor
involves second-order spatial derivatives. When the second-order derivatives are placed in a
covariant formalism, the equations of motion become second order in time,
which leads to runaway solutions and other pathology~\cite{Hiscock:1985zz,Gavassino:2021owo}.  Various approaches
have been adopted to solve this problem. 
One approach,
as initiated  Mueller and  Israel and Stewart (MIS)~\cite{Muller1967, Israel:1976tn,Israel:1979wp},
is to add auxiliary dynamical variables to the system of evolution equations.
There are many variations of this
approach~\cite{Israel:1976tn,Israel:1979wp,Geroch:1990bw,OTTINGER1998433,Denicol:2012cn}, and each
variant involves some additional fields,  which relax on a collisional
timescale $\tau_\pi$  so that the system evolves according to the first-order
hydrodynamics of Landau and Lifshitz at late
times~\cite{Lindblom:1995gp}. Indeed, in a linearized analysis the auxiliary variables lead to gapped ``non-hydrodynamic'' modes, i.e.
modes whose frequency remains finite for $k\rightarrow0$. 
Essentially
all practical simulations of dissipative relativistic fluids have been
based on  variants of the MIS equations.


We will adopt a different approach to hydrodynamics that is truly first order in time and has no non-hydrodynamic modes or additional variables. Recognizing that viscosity controls the diffusion of momentum,  its seems physically reasonable that randomly transferring momentum  between fluid cells,  interspersed
ideal hydrodynamic time steps,  will correctly reproduce the physics of viscosity. 
In trying to clarify this idea and to find compatibility with Metropolis
updates, we found the formulation of hydrodynamic without boosts a clarifying
formalism~\cite{Novak:2019wqg,deBoer:2020xlc,Armas:2020mpr}. 
In particular,  we adopted the
Density Frame formulation of Armas and Jain~\cite{Armas:2020mpr}
(which built upon earlier works~\cite{Novak:2019wqg,deBoer:2020xlc})
and found that it neatly fits with the Metropolis updates used in statistical mechanics, even for relativistic flows in general coordinates. 

In theories without an underlying boost symmetry, hydrodynamics remains valid and is formulated
by writing down the fluxes as spatial gradients of the conserved charge densities\footnote{An example of a fluid without a boost symmetry is a fluid flowing over a fixed surface.}. 
The resulting equations
of motion are first order in time and second order in space. 
If the fluid has Lorentz  symmetry,  then the coefficients of the gradient expansion are related
to each other, but the derivative structure of the equations of motion is unchanged. 
We have previously investigated the stochastic relativistic advection-diffusion equation
in the Density Frame both theoretically and numerically~\cite{Basar:2024qxd}.
Our goal here is to generalize this discussion to the Navier-Stokes system,  and,  in  a companion paper,  to explore the Density Frame numerically for deterministic hydrodynamic flows in $1{+}1{\rm d}$~\cite{NumericsPaper}. The companion paper uses some of the formalism developed here.  

The quickest way to derive the 
equations of motion of the Density Frame is to use lowest-order equations of motion (ideal hydrodynamics) to eliminate time derivatives in the viscous strains. 
The resulting equations of motion are not Lorentz invariant, but are invariant 
under Lorentz transformations followed by a change of hydrodynamic frame.  
This is similar to Heavy Quark Effective Theory in high energy physics,  
which is only covariant to specified order in the $1/m_Q$ expansion~\cite{manohar2000heavy}. 
In essence, each Lorentz observer has  his own hydrodynamic frame.  Indeed,  the Density Frame is a unique hydrodynamic frame where the energy and momentum densities measured on a single spatial slice,  $T^{00}({\bf r})$ and $T^{0i}({\bf r})$,   can be used to reconstruct
the temperature and flow velocity of the fluid.
In the Landau Frame for instance, one would also need the viscous stress $\pi^{0i}$ on the slice,  while in other approaches  at least two spatial slices are needed since the equations of motion are second order in time~\cite{Bemfica:2017wps,Kovtun:2019hdm}. 
The equations of motion in the Density Frame have no additional variables or parameters beyond the shear and bulk viscosities and the equilibrium equation of state.

An outline of the paper is the following. In \Sect{sec:densityframe}  we assemble the  equations of motion in the Density Frame.  While this has been done already in \cite{Armas:2020mpr}, we feel that most readers will benefit from the orthogonal discussion given here.  In \Sect{sec:kinetics} we show how the non-covariant Density Frame 
arises naturally as an approximation scheme for covariant kinetic theory.   
Next in \Sect{sec:metropolis} we 
formulate the stochastic evolution in the Density Frame  as a Metropolis update. For simplicity, in \Sect{sec:stochasticcartesian} we will first describe the algorithm in $2+1$ Cartesian dimensions. 
Briefly, the procedure is the following.  One first takes a step of ideal hydrodynamics. Then a proposal is made for random spatial momentum transfers between fluid cells, which keep the energy fixed.  The proposed momentum transfers are accepted or rejected using the entropy as a statistical weight.  On average this procedure reproduces the mean dissipative stress of the Density Frame, and  the fluctuations inherent in the procedure correctly reproduce the stochastic noise in the system.  
Then in \Sect{sec:bjcoords} we show how the algorithm extends to general coordinates with a specified foliation of space-time. The only complication is that the (spatial) momentum transfers must be  parallel transported from the cell interfaces to the cell centers.

In the present paper, we have not implemented the Metropolis algorithm numerically. In a companion paper~\cite{NumericsPaper}, we conducted a first numerical study of the deterministic Density Frame hydrodynamics for 1+1 dimensional flows. These results are encouraging,  and, in some respects, show that the Density Frame is numerically more robust than other variants of viscous hydrodynamics.
Based on these deterministic numerical results, our previous work on the relativistic stochastic advection-diffusion equation~\cite{Basar:2024qxd}, as well as the strong theoretical foundation of the Density Frame developed here and summarized in \Sect{sec:outlook},  
we hope and expect that the proposed stochastic algorithm will be robust and effective for simulating heavy-ion collisions and other physical systems.

\section{The Density Frame}
\label{sec:densityframe}

First we will write down the equations of motion of  hydrodynamics in the Density Frame and in the process reproduce the form of the viscous tensor given in Eq.~(90) of Armas and Jain~\cite{Armas:2020mpr}.   
The presentation here is markedly different from \cite{Armas:2020mpr}, and we believe that most readers will benefit from the added discussion. 
Further, the shear and bulk parts of the viscous tensor 
are cleanly separated here,    and the form of the Density Frame viscous tensor has an evident mathematical structure,  which  was not clear (to us) in the original work. 

\subsection{Preliminaries} 

Hydrodynamics is  an effective theory for the energy and momentum densities
\st
(T^{00}, T^{0i} ) \equiv (\edense , \pdense^i)\,,
\stp
whose time evolution is given by the conservation laws
\begin{subequations}
  \label{eq:conservation}
\begin{align}
  \partial_t \edense + \partial_i \pdense^i =& 0 \,, \\
  \partial_t \pdense^i + \partial_j T^{ij} =& 0 \,.
\end{align}
\end{subequations}
In order to close the system of equations,  the spatial stress tensor $T^{ij}$ must be specified as a function of  $\edense$ and $\pdense^i$. 

This specification is usually implemented with an intermediate set of parameters,  $\beta_{\mu} = \beta u_{\mu}$, which describe the inverse temperature and four velocity.  
In ideal hydrodynamics the stress tensor  has 
the functional form\footnote{This equation has a small abuse of notation,  which we follow throughout. The stress tensor $\tideal^{\mu\nu}(\beta_{\mu})$ is a function of $\beta_{\mu}=\beta u_{\mu}$ rather than just the scalar $\beta$. However, the pressure and energy density are only functions of $\beta = \sqrt{-\beta_{\mu} \beta^{\mu} }$.  }
\st
\label{eq:tmunuidealparam}
\tideal^{\mu\nu}(\beta) \equiv \left(\er(\beta) + \pr(\beta)\right)  u^{\mu} u^{\nu} + \pr(\beta) \, \eta^{\mu\nu} \, , 
\stp
where $\pr(\beta)$ and $\er(\beta)$ are the pressure and rest frame energy density of the equilibrium equation of state.
This equation for $\tideal^{\mu\nu}(\beta)$ means  that $\beta_{\mu}$ is determined from the energy and momentum densities,
\begin{subequations}
  \label{eq:densityframedef}
\begin{align}
  \edense =& \tideal^{00}(\beta) \, , \\
  \pdense^i =& \tideal^{0i}(\beta)  \,  , 
\end{align}
\end{subequations}
and subsequently $\beta_{\mu}$ is used to specify the spatial stress,  $T^{ij}  = \tideal^{ij}(\beta)$. In the Density Frame
the algebraic relations in \Eq{eq:densityframedef} define $\beta_{\mu}(x)$ to all orders in the derivative expansion.   

However, the Density Frame $T^{ij}$ receives viscous corrections of order $\partial_{(i}\beta_{j)}$. Anticipating the next sections,  we will state    the form of $T^{ij}$ without justification
\st
\label{eq:tjintro}
T^{ij} = \tideal^{ij}(\beta)  + \Pi^{ij}  \qquad \mbox{where} \qquad \Pi^{ij} \equiv - T\noisekernel^{ijmn}(v)\,  \partial_{(m}\beta_{n)}  \, .
\stp
Here $\kappa^{ijmn}(v)$ is proportional to the viscosities of the system,  and  is  a tensor formed with $v^{i}$,  $\delta^{ij}$,  and speed of sound $c_s^2 = d\pr/d\er$.  $\kappa^{ijmn}(v)$ is  symmetric under interchange of $i\leftrightarrow j$ and $m\leftrightarrow n$ and interchange of the index pairs,  $ij \leftrightarrow mn$;  
its form will ultimately given by \Eqs{eq:shear_kappa_ijmn} and \eqref{eq:bulk_kappa_ijmn} for the shear and bulk tensor respectively. 

Next we will derive the form of entropy production in the Density Frame. 
The thermodynamics of Density Frame is the same as an ideal fluid 
with velocity $v$. 
The entropy and four momentum in a spatial volume $V_0 = \int \dd \Sigma_0$ are%
\footnote{ The entropy  per volume is $S(\edense, \pdense)$  is numerically related to the rest fame entropy density parametrized by $\beta$ and the flow velocity,  $S = s(\beta) u^0$. }
 
\st
 \mathcal S =   V_0 S ,   \qquad P^{\mu} = V_0 T^{0\mu} \, , 
\stp
and micro-canonical equation of state ${\mathcal S}(P, V_0)$ determines  $\beta_{\mu}(P)$
\st
\dd {\mathcal S} = -\beta_{\mu}\,  \dd P^{\mu} +  \pr\,  \beta^0 \dd V_0 \, . 
\stp
The Gibbs-Duhem relation follows from extensivity of the system
\st
  S= - \beta_{\mu} T^{0\mu} + \pr \beta^0 \,  . 
\stp
The Legendre transform of the entropy,   ${\mathcal S} + \beta_{\mu} P^{\mu}$,  is the logarithm of the partition function,  $\ln Z(\beta)$,  which is related to the pressure through the Gibbs-Duhem relation
\st
 \ln Z(\beta)  =  \pr(\beta) \beta^0 V_0  \, . 
\stp
The  derivatives of the partition function determine the mean four momentum for a specified $\beta_{\mu}$
\st
\dd \ln Z  =  P^\mu \dd \beta_{\mu} +  \beta^0 \pr  \, \dd V_0 \, . 
\stp
Using the equations of motion and these thermodynamic relations  it is easy to show  that\footnote{One starts with $\partial_t S = -\beta_{\mu} \partial_t T^{\mu 0}$, and then uses the conservation laws and the Gibbs-Duhem relation. The ideal terms yield $Sv^i$, while $\beta_i \, \partial_j \Pi^{ij}$ yields the remaining terms after integrating by parts. }
\st
\partial_t S +   \partial_j (S v^j - \beta_i \Pi^{ij}) = \partial_{(i} \beta_{j)}  \left[ T\kappa^{ijmn}(v) \right] \partial_{(m} \beta_{n)} \, .
\stp
Thus  the positivity of entropy production forces the matrix  $\kappa^{ijmn}$ (with the rows and columns labeled by the index pairs $ij$ and $mn$)
to be positive semi-definite. 

The functional form of $\kappa^{ijmn}(v)$ is unfamiliar. As we will see it reflects the susceptibilities of a boosted fluid,  which will be recorded here for later use.  
The derivatives of the ideal stress tensor form a generalized susceptibility\footnote{Here we  used a relation between the speed of sound and the specific heat:
\st
c_s^2  = \frac{d\pr}{d\er} = \frac{d\pr/d\beta }{d\er/d\beta} = \frac{\er + \pr}{T C_V}. 
 \stp
}
 \st
 \label{eq:Xmunurho}
 \X^{\mu\nu\rho} \equiv  \frac{\partial \T^{\mu\nu}(\beta) }{\partial \beta_\rho} = \frac{\er +\pr}{\beta} \left[ \frac{1}{c_s^2} u^{\mu} u^{\nu} u^{\rho} + (u^{\mu} \Delta^{\nu\rho}  + u^{\nu} \Delta^{\mu\rho} + u^{\rho} \Delta^{\mu\nu})  \right]  \, , 
 \stp
 and also determine the ideal equations of motion 
 \st
 \partial_{\nu} \T^{\mu\nu}(\beta) =  \X^{\mu\nu\rho}\,  \partial_{\nu} \beta_{\rho} = 0   \, . 
 \stp
 Here we defined  $\Delta^{\mu\nu} \equiv \eta^{\mu\nu}  + u^{\mu} u^{\nu}$ as the spatial projector. 
 The thermodynamic susceptibility,   in a strict sense,  is  $\chi^{\mu\nu} \equiv {\mathcal X}^{0\mu\nu}$ and describes the equilibrium fluctuations of the densities $T^{0\mu}$. 
The symmetry of $\chi^{\mu\nu} $ follows directly from the equilibrium partition function
\st
\chi^{\mu\nu} \equiv  \frac{\partial \T^{0\mu}} {\partial \beta_\nu} = \frac{1}{V_0} \frac{\partial^2 \ln Z(\beta) }{\partial \beta_\mu \partial \beta_\nu } \,  . 
\stp
The inverse susceptibility determines the fluctuations of the corresponding Lagrange multipliers $\beta_{\mu}$ and reads 
\begin{align}
  \label{eq:inversesusc}
  \chi^{-1}_{\mu\nu} = \frac{\partial \beta_{\mu}  }{\partial T^{0\nu} }  =& \frac{\beta}{(\er + \pr)\gamma }  \left[  \frac{c_s^2}{1 - c_s^2 v^2} \,    
    \left(u_\mu + \frac{1}{\gamma} \Delta^{0}_{\mu}\right)
    \left(u_\nu + \frac{1}{\gamma} \Delta^{0}_{\nu}\right)
  +    \Delta_{\mu\nu}  \right]   \, .
\end{align}

The factor $c_s^2/(1-c_s^2v^2)$ has a simple interpretation. Indeed, the adiabat is defined by lines of constant $M^i/S$,  and the first term in \eqref{eq:inversesusc} reflects the derivatives of pressure on the adiabat
\st
\left( \frac{d\pr}{d\edense} \right)_{M^i/S} =   \frac{1}{\gamma^2}  \, \frac{c_s^2 }{1 - c_s^2 v^2 } \, . 
\stp
This pressure derivative controls the speed of sound waves propagating transverse to the fluid flow.

We have recorded the thermodynamic derivatives of a boosted fluid because they are needed to evaluate the viscous stress tensor in the Density Frame, \Eq{eq:tjintro}.    In the next sections  we will derive  this tensor by making a change of frame from the covariant Landau Frame. 


\subsection{Hydrodynamic frames}
\label{sec:hydroframes}

Consider hydrodynamics in a general fluid frame. 
The full energy-momentum tensor $T^{\mu\nu}$ is decomposed
into an ideal stress tensor plus viscous corrections
\st
\label{eq:Tmunudecomp}
  T^{\mu\nu}(x) = \T^{\mu\nu}(\beta)  +  \Pi^{\mu\nu}(\beta,\partial \beta)\,.
\stp
However, the decomposition of the stress  into its ideal and viscous 
pieces is not unique. 
Indeed, the ideal stress tensor 
is determined by four parameters $\beta_{\mu}(x) = \beta(x) u_{\mu}(x) $,  whose precise definition specifies the hydrodynamic frame. 
The total stress tensor is independent of
these intermediate parameters.
The constitutive relations provide an
approximate expression for $\Pi^{\mu\nu}(\beta, \partial\beta)$ in terms of the derivatives of
$\beta_{\mu}(x)$ and starts  at  order $\partial \beta$.


Now consider a new frame labeled by $\ubeta_{\nu}(x)$ and 
 $\underline{\Pi}^{\mu\nu}$
\[
T^{\mu\nu}(x) = \T^{\mu\nu}(\ubeta) + \underline{\Pi}^{\mu\nu}(\ubeta, \partial \ubeta)  \, , 
\]
and its relation  to our original frame,  \Eq{eq:Tmunudecomp}.
If $\ubeta_{\rho}$ is redefined  by an amount  of order $\partial \beta$
\begin{align}
    \ubeta_{\rho} = \beta_{\rho} + \delta \beta_{\rho}
\end{align}
the total stress tensor is unchanged.
This means that $\Pi^{\mu\nu}$  and $\underline{\Pi}^{\mu\nu}$ are related by
 \st
 \label{eq:framechange1}
  \Pi^{\mu\nu}(\beta, \partial\beta) = \underline{\Pi}^{\mu\nu}(\ubeta, \partial \ubeta) + \Delta \Pi^{\mu\nu}(\beta, \delta \beta)  \, , 
 \stp
 where  
 \st
  \label{eq:framechange2}
  \Delta \Pi^{\mu\nu}(\beta, \delta \beta) \simeq  \frac{\partial \T^{\mu\nu}}{\partial \beta_{\rho} }  \delta \beta_\rho  = \X^{\mu\nu\rho}\,  \delta\beta_\rho \, ,
%
%
 \stp
 where we have recalled the susceptibility tensor of \eqref{eq:Xmunurho}.
 Since $\underline{\Pi}$ is already first order in derivatives,  we may at first order neglect the differences between $\ubeta$ and $\beta$  when evaluating  $\underline{\Pi}^{\mu\nu}$,   leading to
 \st
 \label{eq:framechange3}
  \Pi^{\mu\nu}(\beta, \partial\beta) \simeq  \underline{\Pi}^{\mu\nu}(\beta, \partial \beta)  + \Delta \Pi^{\mu\nu}(\beta, \delta \beta)  \, . 
 \stp

Certain combinations of $\Pi^{\mu\nu}(\beta)$ are invariant under the reparametrization  of $\beta_{\mu}$.
 Indeed, defining the scalar and tensor projection operators 
 \st
 \label{eq:def_projectors}
 P_{\mu\nu} \equiv-c_s^2 u_{\mu}  u_{\nu} + \frac{1}{d} \Delta_{\mu\nu}\,,    \qquad \mbox{and}  
   \qquad \mathring{P}_{\mu\nu}^{\rho \sigma}  \equiv \Delta^{\rho}_{(\mu} \Delta^{\sigma}_{\nu)} - \frac{1}{d} \Delta^{\rho\sigma}\Delta_{\mu\nu} \,,
   \stp
   we see that 
 \st
 P_{\mu\nu} \,\Delta \Pi^{\mu\nu} =0\,,   \qquad \mbox{and} \qquad \mathring{P}_{\mu\nu}^{\rho \sigma} \,  \Delta\Pi^{\mu\nu}   = 0\,, 
 \stp
 after examining the form of  $\X^{\mu\nu\rho}$. 
 Thus, we define the frame invariant bulk scalar and shear tensors:
 \st
 \Pi_\zeta = P_{\mu\nu} \, \Pi^{\mu\nu}(\beta)  \,, \qquad \qquad \Pi_\eta^{\rho\sigma}  = \mathring{P}_{\mu\nu}^{\rho \sigma} \, \Pi^{\mu\nu}(\beta)\,.
 \stp
 Strictly speaking $P_{\mu\nu}$  is not a projection operator. However, $\bar{P}^{\mu\nu}_{\rho\sigma} \equiv \Delta^{\mu\nu} P_{\rho\sigma}$
 can play this role,  leading to the  algebra of projections
 \st
   \mathring{P}^{\mu\nu}_{\alpha\beta} \, \mathring{P}^{\alpha\beta}_{\rho\sigma} =  \mathring{P}^{\mu\nu}_{\rho\sigma} \, ,  \qquad  
   \bar{P}^{\mu\nu}_{\alpha\beta} \,  \bar{P}^{\alpha\beta}_{\rho\sigma} =  \bar{P}^{\mu\nu}_{\rho\sigma} \, ,  \qquad  
   \bar{P}^{\mu\nu}_{\alpha\beta} \, \mathring{P}^{\alpha\beta}_{\rho\sigma} =  0 \, .
   \stp
 \subsubsection{Landau Frame} 
\label{sec:landaufdef}
 In the Landau Frame, the parameters $\beta_{\mu}$ are chosen such that
 \st
 \beta_{\mu} T^{\mu\nu} = -e(\beta) \beta^{\nu} \,,
 \qquad \mbox{or} \qquad 
 \beta_\mu \Pi^{\mu\nu}(\beta) = 0  \,.
 \stp
 As always, the  viscous stress $\Pi^{\mu\nu}$ is expanded in strains $\partial_{(\mu} \beta_{\nu)}$. However, not all of the gradients in
 this set are independent,  since  the four ideal equations of motion ${\mathcal X}^{\mu\nu\rho}\partial_{\nu} \beta_{\rho }=0$ can be used to express temporal gradients  $u^{\mu} \partial_{\mu} \beta_{\nu} $  in terms of spatial ones $\Delta_{\mu}^{\rho}  \partial_\rho \beta_{\nu}$  to the same order of accuracy. With these considerations in mind, the Landau Frame stress tensor is written as~\cite{landau2013fluid}
 \begin{subequations}
\label{eq:Landaustress} 
\st
\underline{\Pi}^{\mu\nu}=  - T K^{\mu\nu\rho\sigma }  \; \partial_{(\rho} \beta_{\sigma)}\,,
\stp
where
\st
K^{\mu\nu\rho\sigma}  =  
\left[ 2\eta \left(\Delta^{(\mu\rho} \Delta^{\nu)\sigma }  - \frac{1}{d} \Delta^{\mu\nu} \Delta^{\rho\sigma} \right) + \zeta \Delta^{\mu\nu} \Delta^{\rho\sigma} \right] \equiv K_{\eta}^{\mu\nu\rho\sigma} + K_\zeta^{\mu\nu\rho\sigma}. 
\stp
\end{subequations}
The two terms in \eq{eq:Landaustress}  determine the 
frame invariant shear tensor $\Pi^{\mu\nu}_\eta$  
and bulk scalar $\Pi^{\mu\nu}_\zeta$.

\subsubsection{Density Frame}
\label{sec:dfdef}

In the Density Frame, the parameters $\beta_{\mu}$ are chosen such that
\st
n_{\mu} T^{\mu\nu} = n_{\mu} \T^{\mu\nu}(\beta)  \,,   \qquad \mbox{or} \qquad n_{\mu} \Pi^{\mu\nu}(\beta) = 0\,,
\stp
where $n^{\mu} = (1, 0, 0, 0)$ notates the lab frame\footnote{See also section~\ref{sec:bjcoords} and appendix~\ref{sec:generalcoords} where the discussion is generalized to general foliation of space time.}. 
With 
this choice,  the energy per volume $T^{00} = \edense$ and momentum  per volume $M^i = T^{0i}$  determine the  temperature and flow velocity $\beta_{\mu}$. In all other hydrodynamic frames this data is insufficient to determine $\beta_{\mu}$. $\Pi^{\mu\nu}(\beta)$ is expanded in the strains $\partial_{\mu} \beta_{\nu}$. 
However, using the ideal equations of motion  we can rewrite the temporal derivatives $\partial_t \beta_\mu$ appearing in the gradient expansion in terms of spatial derivatives,  $\partial_{(i} \beta_{j)}$.    
Thus, the strains are written as
\st
\Pi^{ij}  = -T\kappa^{ijmn} \partial_{(m}\beta_{n)}\,,    \qquad \Pi^{00} = \Pi^{0i} = 0 \,.
\label{eq:densityframeviscoustensor}
\stp
In the next section, we will write down  the explicit form of $\kappa^{ijmn}$ and identify the 
frame invariants  associated with the shear and bulk viscous tensors.

\subsection{The Density Frame from the Landau Frame} 
\label{sec:shear_density_frame}
In this section we will  derive the Density Frame stress tensor from the Landau Frame or any other frame  where the stress takes the form of \eqref{eq:Landaustress}.
Our first task is 
to rewrite the strains $\partial_{(\mu}\beta_{\nu)}$ in terms of 
the spatial strains $\partial_{(i}\beta_{j)}$, using the lowest order equations of motion to replace the time derivatives with the spatial derivatives.
The next step is to use the frame transformation rules given in \eqref{eq:framechange1} and \eqref{eq:framechange2}, to zero out the temporal components of the viscous stress, $\Pi^{00} $ and $\Pi^{0i}$.

Turning to our first task, the ideal equations of motion are
\st
 \partial_t \T^{0\mu}(\beta) + \partial_j \T^{j\mu}(\beta) = 0 \, , 
\stp
%
which can be written in terms of $\beta_{\nu}$ using the susceptibility matrix $\chi^{\mu\nu}$
 \begin{align}
   \label{eq:beforesymeom}
   \chi^{\mu\nu} \partial_t \beta_\nu  
 =&-   \frac{\partial \T^{j\mu} }{\partial \beta_\rho } \partial_j \beta_\rho   \,.
 \end{align}
The matrices entering in this expression are symmetric, 
which reflects an integrability constraint following from hydrostatic equilibrium and extensivity~\cite{Jensen:2012jh}. 
Indeed,  symmetry of these matrices can also be seen from the explicit
expression for $\X^{\mu\nu\rho}$ in \eqref{eq:Xmunurho},  which is totally symmetric in $\mu \nu \rho$.
Using the symmetry to exchange $\rho$ and $\mu$ in the RHS of \Eq{eq:beforesymeom} and multiplying by the inverse susceptibility $\chi^{-1}_{\sigma \mu} = \partial \beta_\mu / \partial \T^{0\sigma}  $
leads to 
\begin{equation}
 \partial_t \beta_{\sigma} =  - \frac{\partial \T^{j \rho }}{\partial \T^{0\sigma}} \partial_j \beta_{\rho }\,.
\end{equation}
Now using the symmetry of the stress tensor $T^{i0}=T^{0i}$ 
(which is a consequence of relativistic covariance) we find finally
\st
  \partial_{(\mu} \beta_{\nu) }  \simeq 
  \kappa_{\mu\nu}^{ij} 
   \; \partial_{(i} \beta_{j)} \,,
   \qquad \kappa_{\mu\nu}^{ij}  \equiv
  \left( \delta^{i}_{(\mu} \delta^{j}_{\nu)}  - \frac{\partial \T^{ij}}{\partial \T^{0\rho} } \delta^{0}_{(\mu } \delta^{\rho}_{\nu)} \right)  \, . 
 \label{eq:Landau_strains}
\stp
 The transformation matrix  $\kappa_{\mu\nu}^{ij}$ will be  evaluated in explicit form shortly.

 Examining the expression for the $\underline{\Pi}^{\mu\nu}$ in the Landau Frame in \Eq{eq:Landaustress},   
 we see that  we have completed half of our task -- we have expressed  the strains  $\partial_{(\mu} \beta_{\nu)}$ in terms of the spatial data $\partial_{(i}\beta_{j)}$.  The next half is 
 to make a frame change,   using the rules given in \eqref{eq:framechange1} and \eqref{eq:framechange2} to relate the Landau Frame  
 to the Density Frame. This frame change zeros out the  temporal components  of the viscous stress in the Landau Frame
$\underline{\Pi}^{0\nu}$ yielding the Density Frame $\Pi^{ij}$,  where  $\Pi^{00}=\Pi^{0i} = 0$.  
From \eq{eq:framechange2},  we need to choose the frame shift $\delta \beta_{\mu}$ so that    
 \st
 -\chi^{\nu \rho} \delta\beta_{\rho}     = \underline{\Pi}^{0\nu}\,, \qquad \delta\beta _\rho = -\chi^{-1}_{\rho \nu} \, \underline{\Pi}^{0\nu}\,.
 \label{eq:deltabp}
 \stp
 With this choice the temporal components of $\Pi^{\mu\nu}$  are zero by construction, 
 while the spatial components are determined by the Landau Frame stress
 \st
 \Pi^{ij} =
 \left( \delta^{i}_{(\mu} \delta^{j}_{\nu)}  - \frac{\partial \T^{ij}}{\partial \T^{0\rho} } \delta^{0}_{(\mu } \delta^{\rho}_{\nu)} \right) \underline \Pi^{\mu\nu}   \, .
 \stp
 The frame change can also be neatly rewritten using the transformation matrix given in \Eq{eq:Landau_strains} 
 \st
 \Pi^{ij} = \kappa^{ij}_{\mu\nu}  \, \underline{\Pi}^{\mu\nu}   \, . 
 \stp
 Thus we have completed our task of expressing $\kappa^{ijmn}$ in \eqref{eq:densityframeviscoustensor} in 
 terms of $K^{\mu\nu\rho \sigma}$ in \eqref{eq:Landaustress}  
 \st
 \label{eq:kappaijmntrans}
 \kappa^{ijmn} =  \kappa^{ij}_{\mu\nu}\,  K^{\mu\nu\rho \sigma}  \kappa_{\rho\sigma}^{mn}  \, .
 \stp
 In the following section the transformation matrix  $\kappa_{\mu\nu}^{ij}$ will be evaluated in the explicit form, and the noise matrix $\kappa^{ijmn}$ will be evaluated. 
 \subsection{Explicit evaluation of the viscous stress tensor in the Density Frame}
 The goal of this section is to determine the viscous stress tensor given in \Eq{eq:kappaijmntrans} using the transformation matrix defined in \Eq{eq:Landau_strains}. 
Evaluating the relevant derivatives involves differentiating  the ideal stress  tensor
%
\st
  \T^{ij}(\edense, \pdense) =  \frac{\pdense^i \pdense^j  }{\edense + \mathcal P(\edense, \pdense) } + \mathcal P(\edense, \pdense) \delta^{ij} \,,
\stp
with respect to $\edense$ and $\pdense^i$. The algebra is straightforward.  It is beneficial 
to introduce a number of  algebraic structures associated with the comoving  coordinates to make this algebra transparent. 

Comoving coordinates are defined relative to Cartesian-Minkowski coordinates, 
\st
  t  \equiv  t \,, \qquad 
  y^i \equiv  x^i - v^i t \,.
  \stp
The  coordinate differentials  in the comoving frame are
\st
\dd y^i \equiv \el^i_{\mu} \dd x^{\mu }\, ,   \qquad \mbox{with}  \qquad \el^{i}_{\mu} = \delta^{i}_{\mu} - v^i \delta^0_{\mu} \, . 
\stp
We  note that $\el_{\mu}^i  u^{\mu} = 0$. 
The dot product between these differentials $\Lh^{ij} \equiv \dd y^i \cdot \dd y^j$ acts as an inverse metric in the comoving frame
\st
\label{eq:cometric}
\Lh^{ij}  \equiv  \el^i_{\mu} \el^j_{\nu} \, \eta^{\mu\nu}   =  \delta^{ij} - v^i v^j\,.
\stp
Straightforward differentiation shows that these objects appear naturally in the transformation matrix
\st
\kappa_{\mu\nu}^{ij} =   \el^i_{(\mu} \el^j_{\nu)} - \Lh^{ij} \frac{\partial \mathcal P }{\partial \T^{0\rho}} \delta^{0}_{(\mu } \delta^{\rho}_{\nu)}\,.
\label{eq:KappaIJStrain1}
\stp
Using the susceptibilities in \eqref{eq:inversesusc},  the pressure derivatives read
\begin{align}
  \label{eq:pressure_derivs}
  \frac{\partial \mathcal P }{\partial \T^{0\rho }} \delta^{0}_{(\mu } \delta^{\rho}_{\nu) }  
=& -\frac{c_s^2}{1-v^2 c_s^2} \left[ \frac{1}{\gamma^2}  \Delta^0_{\mu}  \Delta^0_{\nu} - 
u_\mu u_\nu \right]\,.
\end{align}

Now let us use the scalar projection operator, repeated here for convenience
\st
 P_{\mu\nu} \equiv -c_s^2 u_\mu u_\nu + \frac{1}{d} \Delta_{\mu\nu}\,,
 \label{eq:scalar_operator}  
\stp
to decompose $\kappa^{ij}_{\mu\nu}$  into its shear and bulk pieces.  
The part of $\Pi^{ij}$ that reflects the shear should vanish
when contracted with the projector, i.e. $P_{ij} \Pi^{ij}_\eta=0$.
By analyzing which terms in \eqref{eq:KappaIJStrain1} vanish when contracted 
with $P_{ij}$, we find the  frame transformation matrix  can be 
decomposed as follows
\begin{subequations}
\label{eq:kappadef}
\begin{align}
  \kappa^{ij}_{\mu\nu}  =& 
  \left( \el^i_{(\mu} \el^j_{\nu)}  -  \frac{\Lh^{ij} }{\llangle  P\Lh \rrangle }  (\el P\el)_{\mu\nu } \right)  +   \frac{\Lh^{ij}}{\llangle  P \Lh \rrangle } P_{\mu\nu} \label{kappadefa} \, ,  \\
  \equiv& \mathring{\kappa}_{\mu\nu}^{ij} + \bar\kappa_{\mu\nu}^{ij}  \, .  \label{eq:kappadefb}
\end{align}
\end{subequations}
In \Eq{eq:kappadef}  we are using a compact matrix notation  
\st
(\el P\el)_{\mu\nu}  \equiv  \el^{i}_{\mu} P_{ij} \el^{j}_{\nu}  = - \frac{c_s^2 }{\gamma^2}  \Delta^{0}_\mu \Delta^0_\nu + \frac{1}{d} \Delta_{\mu\nu}  \, , 
\stp
and using  angular brackets to denote the trace 
\st
  \label{eq:PL}
\llangle  P \Lh \rrangle \equiv  P_{ij} \Lh^{ji} = 1- c_s^2 v^2 \,.
\stp

Now when we apply the transformation matrix  
 \st
           \Pi^{ij} =
	   \left(\mathring{\kappa}_{\mu\nu}^{ij} + \bar\kappa_{\mu\nu}^{ij}  \right) \underline{\Pi}^{\mu\nu} = 
	   \mathring{\kappa}_{\mu\nu}^{ij} \, \Pi_\eta^{\mu\nu} 
	    + \bar{\kappa}_{\mu\nu}^{ij} \, \Pi_\zeta^{\mu\nu}  \, , 
 \stp
we see that first term  from \eqref{eq:kappadefb} projects out the frame invariant shear tensor $\Pi_\eta^{\mu\nu}$, while the second term projects out the frame invariant bulk scalar $\Pi_\zeta^{\mu\nu}$.
 Multiple identities,  
 such as
 \st
   \mathring{\kappa }_{\mu\nu}^{ij}
    = \mathring{\kappa }_{\rho\sigma}^{ij}
    \mathring{P}^{\rho\sigma}_{\mu\nu}  \,,
\qquad
   \bar{\kappa}_{\mu\nu}^{ij}
    = \bar{\kappa }_{\rho\sigma}^{ij}
    \bar{P}^{\rho\sigma}_{\mu\nu}  \,,
    \qquad
     \bar{\kappa }_{\rho\sigma}^{ij}
    \mathring{P}^{\rho\sigma}_{\mu\nu}  
    =
     \mathring{\kappa }_{\rho\sigma}^{ij}
    \bar{P}^{\rho\sigma}_{\mu\nu}    = 0 \,,
    \stp
 enforce the consistency  
 of this decomposition.
Further,  these relations  can be ``undone" by applying 
the projectors  to the $ij$ indices
\st
\label{eq:undone}
\mathring{P}^{\mu\nu}_{ij} \,  \mathring{\kappa}^{ij}_{\rho\sigma} = \mathring{P}^{\mu\nu}_{\rho\sigma}  \,,
\qquad 
\bar{P}^{\mu\nu}_{ij} \, \bar{\kappa}^{ij}_{\rho\sigma} = \bar{P}^{\mu\nu}_{\rho\sigma}\,,
\qquad
\bar{P}^{\mu\nu}_{ij} \,  \mathring{\kappa}^{ij}_{\rho\sigma} 
=
\mathring{P}^{\mu\nu}_{ij} \, \bar{\kappa}^{ij}_{\rho\sigma}  = 0 \,.
\stp
Notably the  shear tensors vanish when contracted with $P_{ij}$ 
\st
P_{ij} \,  \mathring{\kappa}^{ij}_{\mu\nu} =0    \,  , 
\stp
which clearly identifies the shear component after the transformation to the Density Frame.

Finally, we record the viscous stress tensor in the Density Frame
\begin{eqnarray}
    \Pi^{ij} &=& -T \left(\kappa^{ijmn}_\eta + \kappa^{ijmn}_\zeta \right) \partial_{(m} \beta_{n)}\,,
    \label{eq:density_stress_tensor}
\end{eqnarray}
where $\kappa^{ijmn}_\eta$ and $\kappa^{ijmn}_\zeta$ are given by
\begin{eqnarray}
    \kappa^{ijmn} &=& \kappa^{ij}_{\mu\nu}  \left[ 
      2\eta \left(\Delta^{(\mu\rho} \Delta^{\nu)\sigma}  -\frac{1}{d} \Delta^{\mu\nu} \Delta^{\rho\sigma}\right)   
       + \zeta \Delta^{\mu\nu} \Delta^{\rho\sigma}  
       \right]  \kappa^{mn}_{\rho \sigma} \equiv \kappa^{ijmn}_\eta + \kappa^{ijmn}_\zeta\,.
       \label{eq:finalKappa}
       \end{eqnarray}
Using the available identities, 
explicit forms for these tensors can be determined and the shear tensor reads
\begin{eqnarray}
\kappa^{ijmn}_\eta &=& 2\eta \left[  \Lh^{(i m } \Lh^{j) n}  - \frac{\Lh^{ij} }{ \llangle P \Lh  \rrangle } (\Lh P \Lh)^{mn} - \frac{\Lh^{mn}}{\llangle P\Lh \rrangle } (\Lh P \Lh)^{ij}  +  \frac{\llangle P \Lh P \Lh\rrangle }{\llangle P\Lh \rrangle^2 } \Lh^{ij} \Lh^{mn}  \right]\,,
\label{eq:shear_kappa_ijmn}
\end{eqnarray}
where 
\begin{subequations}
\begin{align}
  (\Lh P \Lh)^{ij} = \Lh^{im}P_{mn} \Lh^{nj} =&  - \frac{c_s^2 v^i v^j}{\gamma^2}  + \frac{\Lh^{ij}}{d} \,,\\
  \llangle P \Lh P\Lh \rrangle = P_{ji} (\Lh P \Lh)^{ij} =& 
     \frac{(d-1)}{d} (c_s^2 v^2)^2   + \frac{1}{d} (1- c_s^2 v^2)^2\,.
\end{align}
\end{subequations}
Here the metric $\Lh^{ij}$ is defined in \eqref{eq:cometric},  the scalar projector $P_{ij}$ is given in \eqref{eq:scalar_operator},  and its trace $\llangle P\Lh\rrangle$ is notated with angular brackets and evaluated in \eqref{eq:PL}. 
One can easily verify that 
\st
     P_{ij} \, \kappa_{\eta}^{ijmn} = P_{mn} \, \kappa_\eta^{ijmn} = 0  \, , 
\stp
as should be the case for the shear tensor. 
The bulk tensor is similar and evaluates to
\st
  \kappa^{ijmn}_\zeta = \zeta \frac{\Lh^{ij} \Lh^{mn} }{\llangle P \Lh \rrangle^2 }\,.
  \label{eq:bulk_kappa_ijmn}
  \stp
We can see the relationship between the bulk tensor in the Landau and Density Frames by applying the projection operators leading to 
\st
\bar{P}_{ij}^{\mu\nu} \bar{P}^{\rho\sigma}_{mn}\,  \kappa_{\zeta}^{ijmn}  = \zeta \Delta^{\mu\nu} \Delta^{\rho\sigma}  = K_\zeta^{\mu\nu\rho \sigma}\,.
\stp
Finally, we have checked that results for the shear and bulk tensors presented are consistent with the (rather different) tensor decomposition of Eq. (90) of \cite{Armas:2020mpr}. 

Having analyzed the change of frames between the Landau and Density Frames, we will turn to relativistic kinetics and show how the unfamiliar and non-covariant form of the Density Frame constitutive relation follows from a covariant microscopic theory.

\section{The Density Frame from relativistic kinetics} 
\label{sec:kinetics}

In this section we will show how the Density Frame arises naturally
within relativistic kinetic theory.   Our goal  is to show 
how the viscous stress tensor of the non-covariant Density Frame is carried 
by the relativistic  constituents in covariant kinetics. 
At a practical level of simulating heavy ion collisions,  we wish 
to show  how the first viscous correction $\delta f$ in the Density Frame
is related to the commonly used Landau-frame $\delta f$.

\subsection{Preliminaries}
Consider a bosonic system close to an equilibrium state parametrized by $\beta_{\nu}(x)$. 
The phase space distribution function is decomposed into an equilibrium 
distribution plus a viscous correction 
\st
f(x,p) = f_0(\beta, p) + \delta f(\beta, \delta \beta, p)\,,
\stp
where the dependence on the spacetime coordinates  is through $\beta_{\nu}(x)$. 
Here the equilibrium distribution  function is given by 
\st
\label{eq:f0}
f_0(\beta, p) =  \frac{1}{e^{\beta E_p} - 1}   \, , 
\stp
where $\beta E_p = -p^\mu \beta_\mu$,  and
$\delta f$ corrects this distribution order by order in gradients $\partial \beta(x)$.  
As with the total stress tensor discussed in \Sect{sec:densityframe}, the total phase space distribution $f(x,p)$ is independent of the intermediate parameters  ${\beta}_\nu(x)$. 
In a new frame with $\ubeta_{\mu} = \beta_{\mu} + \delta \beta_{\mu}$,  the  phase space density is unchanged,  but is re-parameterized by $\ubeta$ and $\underline{\delta f}$
\st
f(x,p) = f_0(\ubeta, p) + \delta \underline{f}(\ubeta,  \delta\ubeta, p).
\stp
To relate the two hydrodynamic frames,  we follow \Sect{sec:densityframe}  and expand in $\delta \beta$,  neglecting the differences between $\beta$ and $\ubeta$ in $\delta f$,  which is already first order. Thus 
\st
\label{eq:dfframechange}
  \delta f(\beta, \delta \beta, p) \simeq \delta\underline{f}(\beta, \partial \beta, p)  +  \Delta f(\beta, \delta\beta, p)\,,
\stp
where
\st
\Delta  f =   \frac{\partial f_0}{\partial \beta_{\mu}} \,  \delta\beta_{\mu} = f_0 (1 + f_0) \, p^{\mu} \delta \beta_{\mu} \,.
\stp
The shift $\delta \beta_{\mu}$ is adjusted to reproduce the frame conditions. 
It is straightforward to see that integrating $\Delta f$ over momentum to find the stress tensor reproduces the shift,  $\Delta \Pi^{\mu\nu} = \X^{\mu\nu\rho} \delta\beta_\rho$ in \eqref{eq:framechange2},  where 
\st
\label{eq:Chikinetic}
\X^{\mu\nu\rho} =  \int \frac{d^3p}{(2\pi)^3 p^0} f_0 (1 + f_0) p^{\mu} p^{\nu} p^{\rho}  = \frac{\partial \T^{\mu\nu} } {\partial\beta_\rho}   
\stp
in kinetic theory.

The Boltzmann equation determines the time evolution of $f(x,p)$ and is given by
\st
\label{eq:BE1}
p^{\mu} \partial_{\mu} f = - C_p[f] \, , 
\stp
where $C_p$ is the non-linear collision operator~\cite{FoxUhlenbeck2,Arnold:2000dr}, which vanishes in equilibrium, i.e. $C_p[f_0(\beta)] = 0$ for arbitrary $\beta_\mu(x)$. 
The collision operator is linearized around an equilibrium distribution  to determine 
the first viscous correction. 
Defining\footnote{$\chi_p$ is the thermodynamic conjugate of $f(p)$ as can be seen by expanding the entropy for a specific Fourier mode $s_p = (1 + f(p)) \ln (1 + f(p)) - f(p)\ln f(p)$ to quadratic order in $\delta f$~\cite{FoxUhlenbeck2}.} $\chi_p$  from $\delta f(p)$
\st
\delta f(p) \equiv  f_0 (1 + f_0) \,  \chi_p \, , 
\stp
the collision operator close to the equilibrium state takes the form~\cite{FoxUhlenbeck2,Arnold:2000dr}
\st
\label{eq:BE2}
C_p[f_0(\beta) + \delta f] = C_p[f_0(\beta)]  +  {\mathcal C}_{pk} \circ \chi_{k} = {\mathcal C}_{pk} \circ \chi_k \, . 
\stp
Here $\circ$ indicates the invariant integration over repeated indices, $\int d^3k/(2\pi)^3 k^0$, which defines an inner product between two functions. 
${\mathcal  C}_{pk}$ is a linear positive semi-definite symmetric operator with this inner-product.  
We note that ${\mathcal C}_{pk}$ has a zero mode 
\st
{\mathcal C}_{pk} \circ 
(k^{\mu} \delta \beta_{\mu}) = 0\, , 
\stp
since setting $\delta f(k) = f_0 (1 + f_0) \, k^{\mu} \delta \beta_{\nu}$ is merely a shift in equilibrium, i.e. $f_0(\beta) + \delta f = f_0(\beta + \delta\beta)$. The conservation law $\partial_{\nu} T^{\mu \nu} = 0$ for linearized fluctuations around equilibrium follows from \eqref{eq:BE1} after exploiting the symmetry of ${\mathcal C}_{pk}$ and the zero mode.  


To solve for $\delta f$ in an approximate way we substitute $f_0 + \delta f$ into \eqref{eq:BE1} and neglect terms second order in derivatives to arrive at an integral equation for $\chi_k$ 
\st
\label{eq:linearizedBE}
p^{\mu} p^{\nu} f_0 (1 + f_0)\,  \partial_{\mu} \beta_{\nu} =  -\mathcal C_{pk} \circ \chi_k \, . 
\stp
The  solution to the integral equation is not unique: if $\chi_p$ solves \eqref{eq:linearizedBE},  then   so does
$\chi_p +  p^{\mu} \delta \beta_{\mu}$ for an arbitrary $\delta \beta_{\mu}$. 
  As discussed below,  the frame conditions are used to fixed this ambiguity, leading to a unique solution.    

\subsection{The Landau Frame} 
In the Landau Frame we use the ideal equations of motion 
\begin{align}
  u^{\mu} \partial_{\mu} \beta =& \beta c_s^2 \, \partial_\mu u^{\mu} \, ,  \\
  u^{\mu} \partial_{\mu} u_{\nu} =&  \frac{1}{\beta} \, \Delta_{\nu}^{\mu} \, \partial_\mu \beta \, , 
\end{align}
to eliminate $u^{\mu} \partial_{(\mu} \beta_{\nu)}$ from the strains with the approximation
\st
\partial_{(\mu} \beta_{\nu) } \simeq \left( 
\mathring{P}_{\mu\nu}^{\rho \sigma}
+ P_{\mu\nu} \Delta^{\rho \sigma} 
\right)  \partial_{(\rho} \beta_{\sigma)}  \, .
\stp
Substituting into \eqref{eq:linearizedBE} we arrive 
 at an integral equation for $\chi_p$
\st
p^{\mu} p^{\nu} \left(\mathring{P}_{\mu\nu}^{\rho \sigma} + \bar{P}_{\mu\nu}\Delta^{\rho \sigma}\right)  \partial_{(\rho} \beta_{\sigma)}  =   - \mathcal{C}_{pk} \circ \chi_k  \, .
\stp
Finally,  the Landau Frame condition, 
\st
-u_{\mu} \Pi^{\mu\nu}  = \int \frac{d^3p}{(2\pi)^3 p^0 } E_p \, p^{\nu} \delta f(p) = 0 \, ,   
\stp
fixes the zero mode  ambiguity so that  $\chi_p$ has a unique solution. 

In the Landau Frame the final solution to the integral equation takes the form
\st
\delta f(p) = - T f_0 (1 + f_0)  \left(  \chi_1(E_p)  \, p^{\mu} p^{\nu} \mathring{P}_{\mu\nu}^{\rho \sigma}  + \chi_2(E_p) \Delta^{\rho\sigma} \right) \,  \partial_{(\rho } \beta_{\sigma)}  \, , 
\label{eq:deltaf_Landau}
\stp
where $\chi_1(E_p)$ and $\chi_2(E_p)$ are scalar functions,   
which are  ultimately responsible for the shear and bulk viscosities respectively.  Neglecting the bulk viscosity  for simplicity (and setting $\chi_2$ to  zero),  we determine the viscous tensor  by integrating  $\delta f(p)$ over momenta,
\begin{align}
  \Pi^{\mu\nu} =  \int \frac{d^3p}{(2\pi)^3 } \frac{p^\mu p^\nu}{p^0} \delta f(p) 
  =&
   -T K^{\mu\nu\rho\sigma}_\eta \,   \partial_{(\rho} \beta_{\sigma)} \, . 
\end{align}
Here the tensor structure $K^{\mu\nu\rho\sigma}_\eta$ from \eqref{eq:Landaustress} is determined by the following kinetic integrals
\st
\label{eq:chi1toshear}
  K^{\mu\nu \rho\sigma}_\eta  =
\int \frac{d^3p}{(2\pi)^3p^0} f_0 (1 + f_0) \chi_1(E_p) \, p^{\mu}p^{\nu} p^{\alpha} p^{\beta} \mathring{P}_{\alpha\beta}^{\rho\sigma}  
\, , 
\stp
which relate the shear viscosity to the solution of the integral equation $\chi_1(E_p)$.

The first viscous correction can be written in terms of the frame invariant shear $\Pi^{\mu\nu}_\eta$, 
\st
\label{eq:LFdfwPi}
\delta f(p) =  f_0 (1 + f_0) \frac{1}{2\eta} \chi_1(E_p) \,  p_{\mu} p_{\nu} \Pi^{\mu\nu}_\eta \, , 
\stp
which is the form normally adopted in practical simulations of heavy ion collisions. We will show that  similar form arises in the Density Frame.

\subsection{The Density Frame}

In the Density Frame we use the lowest order equations of motion to eliminate $\partial_t \beta_{\nu}$ from the strains, 
\st
\partial_{(\rho} \beta_{\sigma) } \simeq \kappa_{\rho\sigma}^{mn} \partial_{(m} \beta_{n)} \, ,
\stp
which when  substituted into \Eq{eq:linearizedBE} yields the integral equation in the Density Frame
\st
p^{\alpha} p^{\beta} \, \kappa_{\alpha\beta}^{mn} \partial_{(m} \beta_{n)}   =  - {\mathcal C}_{pk} \circ \chi_k \, . 
\stp

As in the previous section,  we will focus on the shear viscosity and neglect the bulk viscosity for simplicity. Motivated
by the Landau Frame solution to the integral equation,  we write
\st
\label{eq:deltafDF_proposal}
\delta f(p) =  -T f_0 (1 + f_0) \,  \left[ \chi_1(E_p) \, p^{\alpha} p^{\beta}\, \mathring{P}^{\rho\sigma}_{\alpha\beta}   \right]  \, \kappa_{\rho\sigma}^{mn} \partial_{(m} \beta_{n)}   + f_0 (1 + f_0)   \, p^{\mu} \delta\beta_{\mu}  \, , 
\stp
with $\delta \beta_{\mu}$ adjusted to reproduce the frame condition
\st
\Pi^{0\nu} = \int \frac{d^3p}{(2\pi)^3 p^0}  p^0 p^\nu \delta f(p) = 0 \, . 
\stp
This  form clearly satisfies the integral  equation, differing from \eq{eq:deltaf_Landau} by a multiple of the zero mode.
Since without the zero mode the first term in \eqref{eq:deltafDF_proposal} integrates to $\underline{\Pi}^{\mu\nu} = -TK^{\mu\nu\rho\sigma}_\eta \partial_{(\rho} \beta_{\sigma)}$ as in \eqref{eq:chi1toshear}, we choose $\delta \beta_{\mu}$ as in \eqref{eq:deltabp},   leading to  a solution which satisfies the frame conditions 
\st
\label{eq:dfdf}
\delta f(p) =  -T f_0 (1 + f_0) \, \left[ \chi_1 \, p^{\alpha} p^{\beta} \mathring{P}^{\rho\sigma}_{\alpha\beta}    -   p^{\alpha}\, \chi^{-1}_{\alpha\beta} \, K^{0\beta \rho\sigma}_\eta \right]  \, \kappa_{\rho\sigma}^{ij} \partial_{(i} \beta_{j)}    \, . 
\stp

Next we will evaluate the stress, 
\st
\Pi^{ij} = \int \frac{d^3p}{(2\pi)^3 } \frac{p^i p^j}{p^0} \delta f \, , 
\stp
to verify that it has the Density Frame form. 
Using \eq{eq:chi1toshear} and the intermediate result 
\begin{gather}
\int \frac{d^3p}{(2\pi)^3p^0} f_0 (1 + f_0) \, p^{\mu}p^{\nu} p^{\alpha}\,  \Chi^{-1}_{\alpha\beta}  = \frac{\partial \T^{\mu\nu}}{\partial \T^{0\beta} }    \, , 
\end{gather}
following from \eqref{eq:Chikinetic}, it is straightforward to see that   
\st
\Pi^{ij} = - T\, \left[ \kappa^{ij}_{\mu\nu}\,  K^{\mu\nu\rho \sigma}_\eta   \kappa_{\rho\sigma}^{mn} \right]  \partial_{(m} \beta_{n)}  =  - T\kappa^{ijmn}_{\eta}  \, \partial_{(m } \beta_{n)} \,  ,   
\stp
as expected. 

Finally,  we can find a Density Frame equivalent of the Landau result \eqref{eq:LFdfwPi},  by rewriting \eqref{eq:dfdf} using the available identities in \eqref{eq:undone} as
\st
\label{eq:dfdfwPi}
\delta f(p)  =  f_0 (1 + f_0) 
\left( \frac{\chi_1 }{2\eta} \, p_{\mu} p_{\nu} \, \Pi^{\mu\nu}_\eta  -  p^{\mu} \, \chi^{-1}_{\mu\nu}\,  \Pi^{0\nu}_\eta \right) \, . 
\stp
Here we expressed the viscous correction in terms of the frame invariant shear tensor, $\Pi^{\mu\nu}_\eta \equiv \mathring{P}^{\mu\nu}_{ij} \,  \Pi^{ij}$, evaluated entirely from the Density Frame stress $\Pi^{ij}$. We see, as expected, that the viscous correction is the Landau Frame result plus frame shift of  $\delta \beta_{\mu} = -\chi^{-1}_{\mu\nu} \, \Pi^{0\nu}_\eta$.   

To summarize,  we have shown how the constitutive relation of the non-covariant Density Frame in \eqref{eq:tjintro}  follows as an approximation to relativistic kinetics.  
We also have expressed the first viscous correction in the Density Frame in terms of the frame invariant shear tensor which is formed with $\Pi^{ij}$,  \Eq{eq:dfdfwPi}. This  should be compared with the Landau Frame result, \Eq{eq:LFdfwPi}.  In the next section we will study stochastic hydrodynamics in the  Density Frame.

\section{Stochastic hydrodynamics in the Density Frame and the Metropolis algorithm}
\label{sec:metropolis}

In this section we will show that the stochastic hydrodynamics in the Density Frame is naturally 
implemented using the Metropolis algorithm.  Briefly, one takes a step of ideal hydrodynamics. Then, one proposes spatial momentum transfers between fluid cells, with a variance given by the Density Frame noise kernel $\kappa^{ijmn}(v)$. These proposals are accepted or rejected using the entropy as the statistical weight. The procedure produces the dissipative dynamics of stochastic relativistic fluid dynamics. 

\subsection{Cartesian coordinates in 2+1D}
\label{sec:stochasticcartesian}

For simplicity, we will consider Cartesian coordinates  in  
two dimensions, ${\bm r}=(r^1, r^2)= ({\rm x},{\rm y})$, 
and discretize space into finite volume cells of size $d^2r$.
The evolution variables are the lab frame energy density and momentum density ($\edense, \pdense^i$), and the  energy and momentum  
in a fluid cell 
\st
(E, p^i) = d^2r \, (\edense, \pdense^i) \, .
\stp
The equations of motion  are the conservation laws written in \eqref{eq:conservation}, where 
stress tensor is decomposed into an ideal and viscous pieces. 
The  viscous stress  takes the form of an average stress, discussed in
previous sections, plus noise $\xi^{ij}$ 
\st
\label{eq:tvisc2dreproduce}
\tvisc^{ij} =  \bar{\tvisc}^{ij} + \xi^{ij} =  -T \kappa^{ijmn} \partial_{(m} \beta_{n)}  + \xi^{ij}   \, . 
\stp
In order for the stochastic process to 
equilibrate to the probability distribution of the micro-canonical ensemble, 
\st
P[\edense, \pdense]  \propto  \exp\left(\int d^2r \, S(\edense, \pdense) \right)  \delta\left(\int d^2r \, T^{0\mu} - {\mathbb P}^{\mu}\right) \,  , 
\stp
the noise should respect the Fluctuation-Dissipation Theorem (FDT)
\st
\label{eq:noisevariance}
\llangle \xi^{ij}(t,{\bm r})  \xi^{mn}(t',{\bm r'}) \rrangle 
= 2 T \kappa^{ijmn}(v)\,  \delta(t -t') \delta^2({\bm r} - {\bm r}') \, . 
\stp
Here ${\mathbb P}^{\mu}$ is the total four momentum of the micro-canonical ensemble. 

The form of the noise in the Density Frame can also be found by algebraically manipulating  
the Landau Frame.  In the  Landau Frame the viscous tensor  
and noise  are denoted with an underline, 
\st
\underline{\Pi}^{\mu\nu}(x) =    -T K^{\mu\nu\rho\sigma}\,  \partial_{(\rho} \beta_{\sigma)} + \underline{\xi}^{\mu\nu} \, , 
\stp
and variance of the noise  takes the covariant form 
\st
\label{eq:noisecovariant}
\llangle \underline{\xi}^{\mu\nu}(t,{\bm r}) \underline{\xi}^{\rho\sigma}(t', {\bm r}')   \rrangle =  2 T K^{\mu\nu\rho \sigma  }\,  \delta(t -t') \delta^2({\bm r} - {\bm r}') \, . 
\stp
Since the noise in the Density Frame  is related to the noise in the Landau Frame  by the frame change
\st
\label{eq:noiseframechange}
\xi^{ij} = \kappa_{\mu\nu}^{ij} \, \underline{\xi}^{\mu\nu}  \, , 
\stp
it is a simple  matter to show  that 
\st
\llangle \xi^{ij}(t,{\bm r})  \xi^{mn}(t',{\bm r'}) \rrangle 
= 2 T  \left[ \kappa^{ij}_{\mu\nu} K^{\mu\nu\rho \sigma} \kappa_{\rho\sigma}^{mn} \right]  \,  \delta(t -t') \delta^2({\bm r} - {\bm r}') \, , 
\stp
which reproduces  \eqref{eq:noisevariance} after noting \eqref{eq:kappaijmntrans}.  Generating the Landau Frame noise $\underline{\xi}^{\mu\nu}$  covariantly and applying the transformation matrix as in \eqref{eq:noiseframechange}  may be the easiest way to generate the Density Frame noise $\xi^{ij}$.

The proposed algorithm for stochastic hydrodynamics uses operator splitting. First,  the system is evolved with ideal hydrodynamics over a time $\dt$, 
\begin{subequations}
  \label{eq:idealeom2d}
\begin{align}
  \partial_t \edense +  \partial_i \pdense^i  =& 0 \, ,   \\
    \partial_t \pdense^j +  \partial_i \tideal^{ij}(\beta)   =& 0 \, ,
\end{align}
\end{subequations}
and then the system is evolved with a viscous step over the same interval
\begin{subequations}
  \label{eq:visceom}
\begin{align}
  \partial_t \edense  =& 0  \, , \\
    \partial_t \pdense^j +  \partial_i \tvisc^{ij}   =& 0 \, .
\end{align}
\end{subequations}
Next we will show  how the Metropolis algorithm can be used to implement the viscous step,  producing the correct mean and variance of $\tvisc^{ij}$.

\begin{figure}
  \centering
  \begin{minipage}[c]{0.45\textwidth}
  \includegraphics[width=0.9\textwidth]{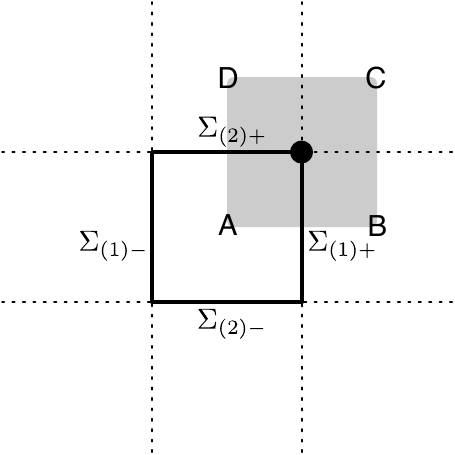}
  \end{minipage}
  \begin{minipage}[c]{0.45\textwidth}
  \includegraphics[width=0.7\textwidth]{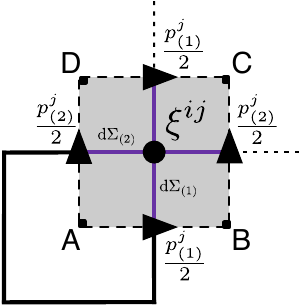}
  \end{minipage}
  \caption{(a) Update of fluid cell $A$ during the Metropolis step. $\dd \Sigma_{\bb1+}$ is
    the area of the right wall of the fluid cell. 
    (b)  Proposed momentum transfers
    when updating cells ABCD due to noise $\xi^{ij}$ living at the corner of the cell (see text). 
    In three dimensions the reader should visualize a cube with cells ${\rm A'B'C'D'}$
    above the ones illustrated. In this case the noise $\xi^{ij}$ is generated at the corner of the cube at the intersection of three planes,  $\dd\Sigma_{\bb1} \ldots \dd\Sigma_{\bbth}$.
  }
  \label{fig:momentumtransfers}
\end{figure}

 To show this, we integrate the viscous equations of motion \eqref{eq:visceom} over the spatial volume of a fluid cell (cell A with volume $\V=d^2r$) and over a time  interval $\dt$  -- see \Fig{fig:momentumtransfers}(a). 
 The change
in the momentum of the cell during the viscous update is determined by the momentum transfers through the walls
of the fluid cell
\st
\label{eq:pupdatecartesian}
(p^i_A)_{t+\dt} - (p^i_A)_t =   
(p_{(1)+}^i  + p_{(1)-}^i) +  
(p_{(2)+}^i  + p_{(2)-}^i)    \, .
\stp
Here, for example, 
\st
 p_{(1)\pm}^i = \mp \dt \, \dd\Sigma_{(1)\pm} \, \tvisc^{1 i } \,,
\stp
is the three-momentum transfer across the $(1)+$ wall with area $\dd \Sigma_{\bb1+} $ (see figure).  There is no change in  the energy 
\st
(E_A)_{t + \dt} -  (E_A)_t = 0  \, . 
\stp

The Metropolis procedure consists of picking a corner of the lattice and updating cells ABCD in a Metropolis accept-reject step (see \Fig{fig:momentumtransfers}(b)), for every group of four cells on the lattice.  Subsequently,  the remaining three corners of cell $A$ can be updated in an analogous way. 

In detail, we envision the noise $\xi^{ij}$  living on the corners of the lattice site (see \Fig{fig:momentumtransfers}(b)), as in our previous study of the advection-diffusion equation~\cite{Basar:2024qxd}. A proposal for the stress $\xi^{ij}$ is generated 
with the Density Frame  variance
\begin{subequations}
  \label{eq:noiseproposal}
\begin{align}
  \llangle \xi^{ij}(t, {\bm r})  \rrangle_{(0)} =& 0  \, ,  \label{eq:variance1} \\ 
  \llangle \xi^{ij}(t, {\bm r}) \xi^{mn}(t', {\bm r}') \rrangle_{(0)} =& \frac{2 T}{\dt \,\V} \, \noisekernel^{ijmn}  
   \delta_{tt'} \delta_{{\bm r} {\bm r}'} \, . \label{eq:variance2}
 \end{align}
 \end{subequations}
Here and below we notate an average over the proposed noise with $\llangle \ldots \rrangle_{(0)}$. 
Associated with the chosen corner are two walls\footnote{As seen in the figure, half of the corner walls $\dd\Sigma_{(1)}$ and $\dd\Sigma_{(2)}$ constitute half of the walls $\dd\Sigma_{(1)+}$ and $\dd\Sigma_{(2)+}$  of the fluid cell A. In three dimensions they would constitute one quarter of $\dd\Sigma_{(1)+}$ and the factors of one half in \eq{eq:pAupdatecart} would be replaced with one quarter.},  $\dd \Sigma_{(1)}$ and $\dd \Sigma_{(2) }$. The proposed momentum
 flux through the corner walls is, for instance,
\st
     \delta p_{(1)}^i = \dt \, \dd\Sigma_{(1) } \, \xi^{1 i } \, ,
\stp
and the proposed update to the  three-momentum for cells $A$ and $B$ are
\begin{subequations}
  \label{eq:pAupdatecart}
\begin{align}
    p_A^i \rightarrow  p_A^i + \delta p_A^i = p_A^i - \half  \delta p_{(1)}^i  -  \half  \delta p_{(2)}^i  \, , \\
    p_B^i \rightarrow  p_B^i + \delta p_B^i = p_B^i + \half  \delta p_{(1)}^i  -  \half  \delta p_{(2)}^i  \, ,
    \end{align}
    \end{subequations}
    with analogous results for neighboring cells $C$ and $D$ - see \Fig{fig:momentumtransfers}(b).

    The change in entropy resulting from the proposed updates is 
\st
\label{eq:dScartesian}
 \Delta \S = \sum_{U \in ABCD} \S[p_U + \delta p_{U} ] - \S[p_U] \simeq -\dt\,\V  \, \xi^{ij} \partial_{(i } \beta_{j) } \,.
\stp
Here we expanded $\S$ to first order in the updates using $\partial \S/\partial p^i = -\beta_i$.   
The derivative in \eqref{eq:dScartesian} is a short hand notation for the discrete difference:
\st
  \partial_x \beta_y \equiv  \frac{1}{2} \left[ (\beta_{B y} - \beta_{A y})/dx  + (\beta_{Cy} - \beta_{Dy})/dx \right]\,.
\stp
In the Metropolis scheme,  the proposed updates are accepted if $\Delta \S > 0$, and are accepted with probability $e^{\Delta \S}$ if $\Delta \S < 0$.  Because of this imbalance, the accepted  proposals $\xi^{ij}$  
develop a mean value, determining the mean stress $\bar{\Pi}^{ij}$.  
The mean value of the proposals $\xi^{ij}$ with the accept-reject step is
\begin{align}
  \label{eq:cartesianmetropolis}
    \btvisc^{ij} = \llangle \xi^{ij} \rrangle  =&   \llangle  \theta(\Delta \S) \xi^{ij} + \theta(-\Delta \S) e^{\Delta \S} \xi^{ij}   \rrangle_{(0)} \,,\\ 
    \simeq & -(\dt\,\V)  \,  \partial_{(m} \beta_{n)} \,  \llangle \theta(-\Delta \S) \,   \xi^{ij}\xi^{mn}   \rrangle_{(0)}\,.
 \end{align}
In passing to the second line we have expanded $e^{\Delta \S} \simeq 1 + \Delta \S$ and used \eq{eq:variance1}.
Owing to the symmetry of the proposal distribution under $\xi^{ij} \rightarrow -\xi^{ij}$,  proposals with $\Delta \S < 0$ occur for exactly half of all the realizations of the noise. Thus the average of $\xi^{ij} \xi^{mn}$ with the   restriction $\Delta \S < 0$ is half that of the unrestricted distribution in  \eqref{eq:noiseproposal}, and the mean stress in the Density Frame reads
 \st
    \btvisc^{ij}   = - T \noisekernel^{ijmn}  \partial_{(m} \beta_{n)}  \,  .
 \stp

 Given the mean stress, the mean momentum transfer to cell $A$ from the first corner  follows from \eqref{eq:pAupdatecart}
 \st
     \llangle \delta p_A^i \rrangle =  -\frac{1}{2} \dt \, \dd\Sigma_{(1)+} \,  \btvisc^{1 i }   -\frac{1}{2} \dt \, \dd\Sigma_{(2)+} \, \btvisc^{2 i} \, ,
 \stp
 which reproduces exactly half of the update in \eq{eq:pupdatecartesian}. The remaining half comes from the update of the lower right corner. 
 A complete viscous update would \begin{enumerate*}[label=(\roman*)] \item loop through the upper right lattice corners,  updating independent groups of four cells associated with this corner as just described;
   \item Then proceed to the remaining corners of cell $A$ and repeat the process one at time.  
     \end{enumerate*}
     To avoid potential bias the order that the corners are updated should be shuffled. 

 When combined with the ideal hydro step, the complete Metropolis algorithm reproduces stochastic viscous hydrodynamics in the Density Frame, with mean viscous stress and noise given in  \eq{eq:tvisc2dreproduce}.
     We followed a similar algorithm when implementing the stochastic advection-diffusion equation numerically in \cite{Basar:2024qxd}.

 \subsection{General Relativity and Bjorken  coordinates} 
 \label{sec:bjcoords}

 The procedure outlined in the previous sections for Cartesian  coordinates 
 readily extends to  general coordinates and to general relativity.  
 Essentially the only change is that the proposed momentum transfers must be parallel transported from
 the cell faces to the cell centers before applying the accept-reject criterion. 
 The definitions of energy and momentum are based on the decomposition of
 the stress tensor according to the fiducial observer  associated with the adopted foliation of space time. 
 Since our primary audience are heavy ion physicists,  we will first present 
 the algorithm in Bjorken coordinates~\cite{Bjorken:1982qr}. Then, in a (somewhat long) appendix, we outline the steps in general coordinates, making full use of the $3+1$ decomposition of spacetime in general relativity.

 In Bjorken coordinates,  $\tau\equiv\sqrt{t^2 -z^2}$ and $\tanh\eta \equiv z/t$,  the metric is
 \st
   \dd s^2 = -\dd\tau^2  + (\dd x^2  + \dd y^2 + \tau^2 \dd \eta^2 )\,.
 \stp
 In comparison to a general coordinate system (see \app{sec:generalcoords}), Bjorken coordinates 
 have a lapse of $N=1$ and a shift of $N^i = 0$, and the spatial metric at a fixed time-slice 
 is flat
 \st
    {}^{(3)}\dd s^2 = \gamma_{ij}\,  \dd r^i \, \dd r^j  =  \dd x^2 + \dd y^2  + \tau^2 \dd z^2 \,,
 \stp
 where $\dd r^i$, with $i=x,y, \eta$, labels the spatial coordinates. The volume element is 
 $\sqrt{g} = \sqrt{\gamma} = \tau$.

 In Bjorken coordinates, the energy and momentum densities and the energy and momentum in a fluid cell are, respectively,
 \st
      (T^{\tau\tau}, T^{\tau i}) \equiv (\edense, \pdense^i) \, , 
      \quad\mbox{with} \quad P^{\mu}=(E, p^i) \equiv \sqrt{\gamma} \, \dd^3r\, T^{\tau \mu}       \, .
 \stp
The conservation laws read
\begin{subequations}
  \label{eq:bjconservationlaws}
\begin{align}
  \partial_\tau (\tau \edense) + \partial_i (\tau \pdense^i)   + \tau^2 T^{\eta\eta} =& 0  \, ,  \\
  \partial_\tau (\tau \pdense^i) +  2 \pdense^\eta \delta^{i}_{\eta}   +    \partial_j (\tau T^{ij})   =& 0   \, .
\end{align}
\end{subequations}
As in previous sections the stress tensor consists of the ideal tensor $\tideal^{ij}(\beta)$ and  the viscous stress, $\tvisc^{ij}$.
The average viscous stress for the Density Frame in Bjorken coordinates is
the same as Cartesian coordinates after replacing the usual derivatives with covariant derivatives: 
\begin{align}
  \label{eq:viscstressbj}
  \btvisc^{ij} = -T \noisekernel^{ijmn} \, \nabla_{(m} \beta_{n)}  =& 
  -T \noisekernel^{ijmn} \left( \partial_{(m} \beta_{n)} +   \beta^{\tau} \Gamma^\tau_{mn} \right)  \,.
\end{align}
Here $\nabla_{i} \beta_j$ notates the covariant derivative. The only non-zero component of the Christoffel connection in this expression is 
\st
     \Gamma^{\tau}_{\eta\eta} = -K_{\eta\eta} = \tau  \, , 
\stp
which, in a more general context, reflects the extrinsic curvature $K_{ij}$ of the foliation of space-time (see \app{sec:generalcoords}). 

In an operator splitting method, the system is first evolved over a time $d\tau$ with ideal hydrodynamics, neglecting the dissipative stress and the noise in \Eq{eq:bjconservationlaws}.   
In a second step  the system is evolved over a time $d\tau$ with the Metropolis algorithm.
The Metropolis step should reproduce the subsequent viscous dynamics 
\begin{subequations}
  \label{eq:viscousevolve}
\begin{align} 
  \partial_\tau (\tau \edense)   + \tau^2 \Pi^{\eta\eta} =& 0 \,, \label{eq:nonconsv} \\
  \partial_\tau (\tau \pdense^i)   + \partial_j (\tau \Pi^{ji})=&  0 \,.
\end{align} 
\end{subequations}
Integrating over a fluid cell and time interval $\Delta \tau$,  we find 
\begin{subequations}
  \label{eq:Bjorkenupdates}
\begin{align}
  (E_A)_{\tau+ \dtau} - (E_A)_{\tau}   =&  - (\dtau\,\V)\,  \Gamma_{\eta\eta}^\tau \,  \Pi^{\eta\eta} \,,  \\
  (p^i_A)_{\tau + \dtau} - (p_A^i)_\tau  =&   
  (p_{(\eta)+}^i + p_{(\eta)-}^i)   
   + (p_{(x)+}^i + p_{(x)-}^i)   
   + (p_{(y)+}^i + p_{(y)-}^i)   \,, 
\end{align}
\end{subequations}
where for instance, 
\st
  p_{(\eta)\pm}^i = \mp\dtau \, \dd\Sigma_{(\eta)\pm} \, \tvisc^{\eta i }\,,
\stp
is the momentum transfer through the $(\eta)$ wall. 
In comparison to the Cartesian case there are two differences. First the energy associated with the adopted foliation,  $\edense\equiv T^{\tau\tau}$,   is not conserved  and  changes due to the viscous stress.  Indeed, this is the viscous contribution to the longitudinal work, which is fundamental to heavy ion phenomenology~\cite{Danielewicz:1984ww}. Second, the mean stress in \eq{eq:viscstressbj}  involves the covariant derivatives $\nabla_{(m} \beta_{n)}$.  We will next show how these differences 
are reproduced by parallel transporting the random momentum transfers between the fluid cells.

\subsubsection{Updates and parallel transport}

The proposed noise is generated at the corner of the fluid cell A,  at the intersection of the planes\footnote{One quarter of the plane $\dd\Sigma_{\bb 1}$  forms one quarter of the plane $\dd\Sigma_{(1)+}$ of
  cell $A$ visualized in \Fig{fig:momentumtransfers}.} $\dd\Sigma_{\bb1} \ldots \dd\Sigma_{\bbth}$  as shown \Fig{fig:momentumtransfers}.
The three-momentum transfer $p^i$ through the surface with normal in the $(1)$ direction is 
\st
\label{eq:pextension}
  p^{\mu}_\bb1 =  p^{i}_\bb1 \ve{i}{\mu}  \,, \qquad    p^{i}_\bb1 =   \dtau \, d\Sigma_{\bb 1 } \, \xi^{1 i} \,.
\stp
Here we have introduce the spatial vectors,  $\ve{i}{\mu} \equiv (0, \delta^{\mu}_{i}) $.
The area of the surface is  
\st
  d\Sigma_{\bb1}  =  \epsilon_{1ij} \, dr^i dr^j\,,
\stp
where the epsilon symbol appropriate to the surface is $\epsilon_{ijk} = \sqrt{\gamma}[ijk]$. 
We also denote, $dr^\mu_{\bb1} =  dr^1 e_1^{\mu}$ \
as the change in spatial coordinate normal to the surface, so that,  for instance,
\st
\label{eq:volume}
 \V \equiv  \dd r^1 d\Sigma_{\bb1 } = \tau \dd^3r\,,
\stp
 is the spatial volume of a fluid cell.
 The noise $\xi^{ij}$ is drawn from the proposed distribution of the Density Frame as in \eq{eq:noiseproposal}.

In the update of cells ABCD and ${\rm A'B'C'D'}$ (visualized in \Fig{fig:momentumtransfers}(b)),  the  proposals for the momentum transfer between cells 
are parallel transported from the cell face to the cell center along the appropriate links. For example, the three momentum transfer, $p^{\mu}_\bb1/4$,  which is transferred to $B$ and removed from $A$ at their common cell interface, is parallel transported along the path connecting $A$ to $B$, leading to the following increments to the four-momentum of the cells
  $A$ and $B$
\begin{align}
  \label{eq:pupdatesparallelAB}
    p^\mu_\bb1 \rightarrow  \delta P^\mu_{B\bb1} = (\delta E_{B\bb1}, \delta p^i_{B\bb1} )\equiv&  \tfrac{1}{4} \left( p^\mu_\bb1  - \tfrac{1}{2} \Gamma^{\mu}_{\rho \sigma} dr^\rho_\bb1  p^\sigma_\bb1   \right) \, , \\
    p^\mu_\bb1 \rightarrow  \delta P^\mu_{A\bb1} = (\delta E_{A\bb1}, \delta p^i_{A\bb1} )\equiv&    - \tfrac{1}{4} \left( p^\mu_\bb1  + \tfrac{1}{2} \Gamma^{\mu}_{\rho \sigma} dr^\rho_\bb1  p^\sigma_\bb1 \right) \, .
 \end{align}
 We have notated the four-momentum increment with a capital letter $\delta P^{\mu}$ because the energy transfer during the proposal 
 is non-zero as a result of parallel transport. Specifically, for a Bjorken 
 expansion the energy increments are 
 \begin{align}
   \delta E_{A(\eta)} = - \tfrac{1}{8} \Gamma^{\tau}_{\eta\eta}\,  dr^\eta \, p^\eta_{(\eta)} \,,
   \end{align}
 although the momentum proposals in \eq{eq:noiseproposal} at the interfaces are purely spatial.
 This is ultimately responsible for the shear stress in \eq{eq:nonconsv} and the non-conservation of $\edense$ in the viscous step. 
 The proposed four momentum updates in each cell as a result of the three-momentum transfers at
 the interfaces are  for example 
 \begin{subequations}
   \label{eq:PAupdatebj}
 \begin{align}
 P_A \rightarrow& P_A + \delta P_A = P_A  + \delta P_{A\bb1}  +   \delta P_{A\bbt}   
 + \delta P_{A\bbth}\,,
     \end{align}
     \end{subequations}
with analogous formulas for ${\rm BCD}$ and ${\rm A'B'C'D'}$ -- see \Fig{fig:momentumtransfers}(b).

The corresponding change in the entropy from these momentum transfers
 \st
 \label{eq:dSbj3d}
\Delta \S  = 
\sum_{U \in {\rm ABCDA'B'C'D'}} \S (P_U + \delta P_U) - \S(P_U)\,,
 \stp
 leads to
 \begin{align}
   \Delta \S  =&  -\, \covD_{\mu} \beta_{\nu}   \, \ve{1}{\mu} \ve{i}{\nu}  \, \dd r^1  p_\bb1^i  
   -\covD_{\mu} \beta_{\nu} \, \ve{2}{\mu} \ve{i}{\nu} \,  \dd r^2  p_\bbt^i  
   - \covD_{\mu} \beta_{\nu}    
    \, \ve{3}{\mu} \ve{i}{\nu} \, \dd r^3  p_\bbth^i 
   \, .
   \end{align}
 In this expression $\nabla_\mu \beta_\nu$ is shorthand for  the discrete approximation 
  \begin{multline}
  \ve{1}{\mu} \ve{j}{\nu} \nabla_{\mu} \beta_{\nu}  \equiv     
          \quarter \left[ \ve{j}{\nu}  (\beta_{B\nu}  - \beta_{A\nu})/\dd r^1  - \half \Gamma_{\rho \sigma}^{\nu}  \ve{1}{\rho} \ve{j}{\sigma} (\beta_{B\nu} + \beta_{A\nu})  \right]  \\
 + \mathrm{(AB \rightarrow CD)} + \mathrm{(AB\rightarrow A'B')} + \mathrm{(AB\rightarrow C'D')}\,.
          \end{multline}
%
    Using the symmetry of the noise $\xi^{ij}=-\xi^{ji}$ and noting \eq{eq:volume},  the change in entropy as a result of the momentum transfers is 
    \st
 \Delta \S = -\dtau \V \, \xi^{ij} \,  \nabla_{(i} \beta_{j)}  \, .
    \stp

 At this point we may just repeat the discussion surrounding \Eq{eq:cartesianmetropolis} and compute the mean stress between fluid cells with the analogous result
\st
    \btvisc^{ij} =   \llangle \xi^{ij} \rrangle = - T \noisekernel^{ijmn} \,  \nabla_{(m} \beta_{n)} \,.
\stp

Given the mean stress,  the mean update of cell $A$ as a result of the Metropolis increments  is
\begin{align}
\llangle \delta E_A  \rrangle  =&  -\frac{1}{8}  (\dtau \V) \, \Gamma_{\eta\eta}^\tau \, \btvisc^{\eta\eta} \, , \\
\llangle \delta p_A^i  \rrangle =&  -\frac{1}{4} \dtau \, \dd\Sigma_{(\eta)} \, \btvisc ^{\eta i }  - \frac{1}{4} \dtau \,  \dd\Sigma_{(x)}\,  \btvisc^{x i }   
 - \frac{1}{4} \dtau \,  \dd\Sigma_{(y)}\,  \btvisc^{y i }   \, .
\end{align}
As in Cartesian coordinates, this reproduces the expected viscous Bjorken dynamics of the Density Frame given in \eq{eq:Bjorkenupdates}, after each of the eight corners of the fluid cell is visited.  

In summary, we have shown how Metropolis updates reproduce the mean viscous stress in the Density Frame,  \Eq{eq:viscstressbj}. The stochastic nature of the algorithm automatically reproduces the noise. 
When these stochastic updates are  complemented  with the symplectic steps of ideal hydrodynamics, the stochastic fluid motion is correctly evolved.

\section{Outlook}
\label{sec:outlook}

In the previous sections  we outlined in detail how the Metropolis algorithm and the Density Frame formulation of viscous hydrodynamics can be combined to form a tool for simulating
stochastic viscous fluids in heavy ion collisions and general relativity. However, we have stopped short of actually simulating the
stochastic dynamics in this work.  Clearly  this is the next step,  and it is a step that is well 
motivated by the current manuscript, our previous theoretical and numerical work on the stochastic advection-diffusion equation~\cite{Basar:2024qxd}, and a companion paper~\cite{NumericsPaper}, where we simulated the deterministic Density Frame dynamics in 1+1 dimensions.

From a theoretical perspective
the Density Frame is attractive, representing relativistic hydrodynamics in its purest form. 
Indeed, the Density Frame is the only formulation of relativistic viscous hydrodynamics that has no additional parameters beyond the shear 
and bulk viscosities and the equation of state.  It combines a symplectic step of ideal hydrodynamics with a viscous step that fits
nicely into the framework of dissipative stochastic processes (see for example~\cite{FoxUhlenbeck,Generic}). For this reason the approach is particularly useful for  simulating dynamical critical phenomena. Indeed,  this paper was inspired by our own simulations of the $O(4)$ critical point in QCD for non-expanding fluids at rest~\cite{Florio:2023kmy}.  Recently, in a pioneering paper,  Chattopadhyay, Ott, Schaefer and Skolkov simulated the real-time dynamics of the liquid-gas critical point in QCD using an approach that is similar to the algorithm suggested here, although again the simulation was limited to non-expanding fluids at rest~\cite{Chattopadhyay:2024jlh}. Given the outlines of our work and \cite{Chattopadhyay:2024jlh},  it should be possible to simulate both the $O(4)$ and liquid-gas critical points in the relativistic and expanding Bjorken geometry  used in heavy ion collisions.
We also hope to investigate the fascinating renormalization group properties of two dimensional stochastic fluids~\cite{Kovtun:2012rj}, which would provide an interesting test of the proposed algorithm.



In summary, this work and a companion paper~\cite{NumericsPaper} strongly motivate a  Metropolis implementation  of the Density Frame description of stochastic relativistic fluids. 
One can  imagine using powerful variants of Metropolis algorithm,  such as Hybrid Monte Carlo~\cite{DUANE1987216},  to simulate the stochastic  evolution of the QGP close to its critical points and in small systems. 

\begin{acknowledgments}
    Valuable discussions with G\"ok\c ce Ba\c sar are gratefully acknowledged.
J.B. and D.T. are supported by the U.S. Department of Energy, Office of Science, Office of Nuclear Physics, grant No. DE-FG-02-88ER40388.  
    R.S. is supported partly by a postdoctoral fellowship of West University of Timișoara, Romania.
\end{acknowledgments}

\appendix
\section{The Metropolis algorithm in general coordinates} 
\label{sec:generalcoords}

 \subsection{Preliminaries}
 \label{sec:prelims}
 The formalism of the Density Frame and the Metropolis updates fit naturally 
 within the $3+1$ split of general relativity. 
 The material presented here is standard and we recommend the review article~\cite{Gourgoulhon:2007ue}. 
 The space time metric is decomposed as follows
 \st
   ds^2 = -N^2 \dd t^2 + \gamma_{ij} (\dd x^i + N^i \dd t) (\dd x^j + N^j \dd t)\,.
 \stp
 In the foliation of space-time,  each slice at fixed coordinate time, $\Sigma_t$,   has 
 coordinates $y^i$. The vector normal to the surface of constant coordinate time 
 is $n^{\mu}$ and the vectors tangent to the surface are $\ve{i}{\mu}$.
 In the preferred coordinate system $x^{\mu} = (t, y^i)$
 \st
 \ve{i}{\mu} = \frac{\partial x^\mu}{\partial y^i} = (0, \delta^{\mu}_i)\,,  \qquad n^{\mu}= \left(\frac{1}{N}, -\frac{N^i}{N} \right) \, . 
 \stp
 The dual basis is chosen such that  $\de{j}{\mu}n^{\mu} = 0$ and that 
 \st
 \de{j}{\mu}
 \ve{i}{\mu} \, 
 = \delta_i^{j} \,. 
 \stp
 In component form
 \st
  \de{i}{\mu} = (N^i , \delta^{i}_{\mu}) \, ,   \qquad    n_{\mu} = (-N, 0, 0, 0) \, .
 \stp
 The spatial projector is
 \st
\gamma^{\mu}_{\sp \nu}\equiv \ve{i}{\mu} \de{i}{\nu} = g^{\mu}_{\sp \nu} + n^{\mu} n_{\nu} \, ,
 \stp
 and is used below to decompose vectors and tensors into their spatial and temporal components according 
 to the fiducial observer of the space-time foliation.

The thermal velocity $\beta^{\mu}=\beta u^{\mu}$ is decomposed as follows
\st
  \beta^{\mu} =  \ve{i}{\mu}\vbeta^i  + \beta^\tau n^{\mu}   \, , 
\stp
where  $\beta^\tau = -n \cdot \beta$ and note $\beta^i \neq \vbeta^i$.
The four velocity  is decomposed as 
\st
    u^{\mu} = \gamma v^{i} \ve{i}{\mu} +  \gamma n^{\mu}  \, , 
\stp
and thus the three velocity recorded by the fiducial observer is 
\st
  v^{i} =   \frac{1}{N} \left( N^i + \frac{u^{i}}{u^0} \right) \, , 
\stp
where $u^i/u^0$ is the coordinate velocity.

The stress tensor is decomposed into its temporal and spatial parts 
\st
\label{eq:Tmunudecomp1}
  T^{\mu\nu} = \edense n^{\mu} n^{\nu}  + 2 \pdense^{i} n^{(\mu} \ve{i}{\nu)} +  \Sstress^{ij} \ve{i}{\mu} \ve{j}{\nu}\,.
\stp 
A four momentum vector with spatial and temporal components is notated with a capitol letter $P^{\mu}$ and is decomposed as follows
\st
  P^{\mu} \equiv -\sqrt{\gamma} \, \dd^3r \, n_{\nu} T^{\mu\nu}  = E n^{\mu}  + p^i \ve{i}{\mu}  \, , 
\stp
note that $P^i \neq p^i$.

Covariant derivatives with the full 4-metric of the manifold $g_{\mu\nu}$ are notated $\covD$ and with traditional semi-colon notation:
\st
  \beta^\mu_{\spm;\nu} \equiv \covD_{\nu} \beta^\mu = \partial_{\nu} \beta^{\mu} + \Gamma_{\rho\nu}^{\mu} \beta^\rho\,.
\stp
Covariant derivatives with respect to the three metric $\gamma_{ij}$ are written with  $\vcovD_j$ and with a $|_j$ notation.
Thus for three vectors  $a \in T_{p}(\Sigma) $
\st
   a^i_{\sp|j} \equiv \vcovD_{j} a^i = \partial_{j} a^{i} + {}^{(3)}\Gamma_{kj}^{i} a^k\,.
\stp
We will also need the Lie derivative for four dimensional vectors and forms 
\st
\DD_{w} v^{\mu} = w^{\rho} \partial_\rho v^\mu - v^{\rho} \partial_{\rho} w^{\mu} \,, \qquad    
\DD_{w} v_{\mu} = w^{\rho} \partial_\rho v_\mu + v_{\rho} \partial_{\mu} w^{\rho}  \,,
\stp
and for three dimensional vectors and forms $a,b \in T_p(\Sigma)$:
\st
\DD_{a} b^i = a^{j} \partial_j b^i - b^{j} \partial_{j} a^{i} \,, \qquad    
\DD_{a} b_i = a^j \partial_j b_i + a_{j} \partial_{i} b^{j}  \,.
\stp
The distinction between the Lie derivative in three and four dimensions  will be clear from  the context. 

Consider a vector that is purely spatial such as the three-momentum, $p^{\mu} \equiv p^{i} \ve{i}{\mu}$.  When this vector is parallel transported from one point to another on the spatial slice of the foliation,  it will not remain spatial.  The extrinsic curvature determines the temporal component after parallel transport 
\st
\label{eq:Kijparalleltransport}
p^{\mu}_{\spm;\nu}  \, \ve{i}{\nu}  =  \ve{j}{\mu}\, p^{j}_{\sp|i}  - n^{\mu} p^{j} K_{ji} \,.
\stp
The extrinsic curvature can be written as
\st
\label{eq:Kijdefgeneral}
  K_{ij} =  -n_{\mu;\nu} \, \ve{i}{\mu}\ve{j}{\nu} = \Gamma_{ij}^{\mu} n_{\mu} = - N \Gamma_{ij}^t \,.
\stp

 \subsection{Covariant conservation laws}

 We use the 3 + 1 decomposition of space-time, and write out the covariant conservation laws $\nabla_{\nu} T^{\mu\nu} = 0$,  with the decomposition of $T^{\mu\nu}$ in  \eqref{eq:Tmunudecomp}.
 The equation of energy conservation reads~\cite{Gourgoulhon:2007ue}
 \begin{align}
   \label{eq:geneconsv}
\left(\frac{\partial}{\partial t}  - N^i \partial_i  \right)  (\sqrt{\gamma} \edense) +   \sqrt{\gamma} \, \vcovD_i (N \pdense^i)  + N \pdense^i  a_i  -   N \sqrt{\gamma}  K_{ij} \Sstress^{ij} = 0  \,,
   \end{align}
   where $a_i = (1/N) \partial_i N$ and
   the momentum conservation equation is
   \st
   \label{eq:genpconsv}
   \left(\frac{\partial}{\partial t}  - \DD_{\sft}  \right)   (\sqrt{\gamma} \pdense^{\;i})  -  2 N \sqrt{\gamma} K^{i}_{\;j} \, \pdense^{\;j}  + N \edense a^i +  \sqrt{\gamma}\, \vcovD_j (N \Sstress^{ij} ) = 0  \,.
   \stp
   The spatial stress  $\Sstress^{ij}$ is decomposed into its  ideal and viscous pieces
   \st
         \Sstress^{ij} = \tideal^{ij}(\beta) + \tvisc^{ij}\,.
   \stp
   In general coordinates the mean stress involves the covariant  derivative  
   \st
         \btvisc^{ij} = -T\kappa^{ijmn} \beta_{m;n}\,,
   \stp
   which can be written  in several informative forms
   \st
     \beta_{i;j} =  \beta_{\mu;\nu} \, \ve{i}{\mu} \ve{j}{\nu} 
     = \vbeta_{i|j} - \beta^\tau K_{ij} 
     = (\DD_\beta g_{\mu\nu})\, \ve{i}{\mu} \ve{j}{\nu} \,.
   \stp
   The second form $\vbeta_{i|j} - \beta^\tau K_{ij}$ expresses the relevant derivatives
   in terms of the spatial data of the foliation and is the most important in this context.  The last form shows that if $\beta^{\mu}$ 
   is a Killing vector  of the metric the viscous strains vanish~\cite{Jensen:2012jh}. 
   In the Density Frame only the spatial components of the thermal-metric Lie derivative are used to parameterize the viscous stress. 

   We again use operator splitting and first evolve  ideal hydrodynamics written in \eq{eq:geneconsv} and \eq{eq:genpconsv} for a time $\dt$,  approximating the stress $\Sstress^{ij}$ with  the ideal stress $\tideal^{ij}(\beta)$. This is followed by a stochastic viscous updates, which should evolve
   \begin{align}
     \frac{\partial}{\partial t}  (\sqrt{\gamma} \edense)  -   N \sqrt{\gamma}  K_{ij} \tvisc^{ij} =& 0  \,, \\
     \frac{\partial}{\partial t}(\sqrt{\gamma} \pdense^{i})   + 
     \sqrt{\gamma}\, \vcovD_j (N\tvisc^{ji} ) =& 0 \,,
     \end{align}
     over a time $\Delta t$ to incorporate the viscous correction.
     Noting that 
     \st
     \sqrt{\gamma}\, \vcovD_j (N \, \tvisc^{ij} )  = \partial_j (N\sqrt{\gamma} \tvisc^{ij} )  + N\sqrt{\gamma}  \, \gammathree_{kj}^i \tvisc^{kj} \,,
     \stp
     we follow the general strategy of flat space and Bjorken coordinates, and integrate the equations of motion over a fluid cell,   leading to the update rule
     for the energy conservation
     \begin{subequations}
       \label{eq:gencoordupdate}
       \st
     (E_A)_{t + \dt} - (E_A)_t  =   (N\dt\,\V)\, K_{ij} \tvisc^{ij} \,,
     \stp
     and momentum conservation
     \begin{align}
       (p_A^i)_{t + \dt} - (p_A^i)_t  =&   
     \left(p_{A\bb1+}^i  + p_{A\bb1-}^i\right)  + \left(p_{A\bbt+}^i  + p_{A\bbt-}^i \right) 
       + \left(p_{A\bbth+}^i  + p_{A\bbth-}^i \right)   \,.
       \end{align} 
     \end{subequations}
     Here,  for example,  
     \begin{align}
       p_{A\bb1 \pm}^i =& \mp N\dt\,\dd\Sigma_{\bb1\pm}\, \tvisc^{i1}  
       -  \half (N\dt\,\V) \, \gammathree^{i}_{j1} \tvisc^{j1}\,,
       \end{align}
       is a momentum increment to cell $A$ resulting from a momentum transfer across the corresponding cell wall $\dd\Sigma_{\bb1\pm}$,   which was subsequently parallel transported  to the center of cell $A$.
       We will briefly describe how the Metropolis algorithm reproduces this dynamics in the next section.

       \subsection{Metropolis dynamics in general coordinates}

       This section parallels \Sect{sec:bjcoords} closely and thus  we will be quite brief. 
       The proposed noise follows \Eq{eq:pextension} and \Eq{eq:noiseproposal} with the following replacements
       and identifications:
       \st
       (\dt \;\; \mbox{or} \;\; \dtau)  \rightarrow N\dt \,, \qquad  \V \equiv \sqrt{\gamma} \, \dd r^1 \dd r^2 \dd r^3 \,.
       \stp
       The momenta in the fluid cells are updated by parallel transporting the momentum 
       transfers from the cell interfaces back to the cell centers. 
For example, the three momentum $p^{\mu}_\bb1$/4 given to $B$ and removed from $A$ at the cell interface is then parallel transported along the link connecting $A$ to $B$ as in \eq{eq:pupdatesparallelAB}.
    However, in general coordinates  the decomposition into energy  and three momentum increments must be generalized:
    \st
    \delta P^\mu_{A\bb1} \equiv  \delta E_{A\bb1}  n^{\mu}  +  \delta p^{i}_{A\bb1} \ve{i}{\mu}  \, . 
    \stp
    This yields the increments 
    \begin{subequations}
    \begin{align}
      \delta E_{A\bb1} =&  \tfrac{1}{8} K_{j1}  \, p_{\bb1}^j \, \dd r^1 \, , \\
      \delta p_{A\bb1}^j =& \tfrac{1}{4} \left(-p_{\bb1}^j  -\half \, \gammathree_{k1}^j\;  p_{\bb1}^k  \, \dd r^1  \right)\,,
    \end{align}
    \end{subequations}
    which reflect the parallel transport rules given in \eq{eq:Kijparalleltransport} and \eq{eq:Kijdefgeneral}.  The complete increment for cell $A$  is
    \begin{align}
      \delta E_A =& \delta E_{A\bb1} + \delta  E_{A\bbt} + \delta E_{A\bbth} \, ,  \\
      \delta p_A^i =& \delta p_{A\bb1}^i + \delta  p_{A\bbt}^i + \delta p_{A\bbth}^i \, . 
         \end{align}
    Using the  thermodynamic derivative
    \st
       \frac{\partial \S}{\partial P^\mu} \delta P^{\mu} = -\beta_{\mu} \, \delta P^{\mu} = \beta^{\tau} \delta E - \vec{\beta}_i\, \delta p^i \, , 
    \stp
    the change in entropy from the updates for cells $\mathrm{ABCDA'B'C'D'}$  is analogous to \eq{eq:dSbj3d} and yields
    \st
       \Delta \S = -\dt\,  \V\,  \xi^{ij} \covD_{(i} \beta_{j)} = -\dt \, \V \, \xi^{ij}
       \left( \vcovD_{(i}\vec{\beta}_{j)}- \beta^\tau K_{ji} \right) \, .
    \stp
    From here it is straightforward to apply the discussion in \eq{eq:cartesianmetropolis} to
    show that the mean stress has the required form
\st
    \btvisc^{ij} =   \llangle \xi^{ij} \rrangle = -T \noisekernel^{ijmn} \,  
       \left(\vcovD_{(m}\vec{\beta}_{n)} - \beta^\tau K_{mn} \right) \, .
\stp
Given the mean stress,  the mean update of cell $A$ as a result of the Metropolis increments  is
\begin{align}
\llangle \delta E_A  \rrangle  =&  \frac{1}{8}  (N \dt\,\V) \, K_{ij} \, \btvisc^{ij} \, , \\
\llangle \delta p_A^i  \rrangle =&  \frac{1}{4} \left[(-N\dt\, \dd\Sigma_{(1)} \, \btvisc ^{1 i }  - \tfrac{1}{2}(N\dt\,\V)\,\gammathree^{i}_{k1} \btvisc^{k1})  + 
  (1\rightarrow 2)
 + (1\rightarrow 3)\right] \, .
%
\end{align}
Finally, after updating all cells in the lattice with the first corner,  and then repeating the process for the additional seven corners of the fluid cells, the Markov chain reproduces the expected viscous dynamics of the Density Frame given in \eq{eq:gencoordupdate}. When the viscous step is combined with the ideal step outlined in \eq{eq:geneconsv} and \eq{eq:genpconsv},  the stochastic viscous fluid is correctly evolved.

\end{document}